\documentclass[12pt]{article}

\usepackage{amsmath}
\usepackage{amssymb}
\usepackage{amsfonts}
\usepackage{latexsym}
\usepackage{color}
\usepackage{graphicx}

\catcode `\@=11 \@addtoreset{equation}{section}
\def\theequation{\arabic{section}.\arabic{equation}}
\catcode `\@=12

\newcommand{\be}{\begin{equation}}
	\newcommand{\en}{\end{equation}}
\newcommand{\bea}{\begin{eqnarray}}
	\newcommand{\ena}{\end{eqnarray}}
\newcommand{\beano}{\begin{eqnarray*}}
	\newcommand{\enano}{\end{eqnarray*}}
\newcommand{\bee}{\begin{enumerate}}
	\newcommand{\ene}{\end{enumerate}}

\newcommand{\B}{{\mathfrak B}}

\newcommand{\mc}{\mathcal}

\newcommand{\D}{{\mc D}}

\newcommand{\Sc}{{\cal S}}

\newcommand{\F}{{\cal F}}

\newcommand{\Bc}{{\cal B}}

\newcommand{\1}{1 \!\! 1}

\newcommand{\Hil}{\mc H}

\catcode `\@=11 \@addtoreset{equation}{section}
\catcode `\@=12

\begin{document}
	
	\thispagestyle{empty}

	\vspace*{2cm}
	
	\begin{center}
		{\Large \bf Bank Deposits as {\em Money Quanta}}   \vspace{2cm}\\
		
		{\large Fabio Bagarello}\\
	Dipartimento di Ingegneria,\\[0pt]
	Universit\`{a} di Palermo, I - 90128 Palermo,\\
	and I.N.F.N., Sezione di Catania\\
	E-mail: fabio.bagarello@unipa.it\\
		\vspace{4mm}
		{\large Biagio Bossone}\\
		International financial consultant,\\
		Senior adviser, Finance, Competitiveness and Innovation, The World Bank\\
		e-mail: biagio.bossone@gmail.com and BBossone@WorldBank.org\\
	
		\vspace{2mm}

	\end{center}
	
	\vspace*{2cm}
	
	\begin{abstract}
		\noindent According to the Accounting View of Money (AVM), the money issued by commercial banks in the form of demand deposits features a hybrid nature, since deposits can be shown to consist of a share of deposits bearing the characteristics of debt (debt-deposits) and a share of deposits bearing the characteristics of equity (equity-deposits), in a mix that depends on factors that relate to the issuing banks and the environment where they operate and interact, which may change over time. Following this important finding of the AVM, it is only consequential to associate the hybrid nature of bank deposits with the dual nature of the objects which is typical in quantum physics, and to investigate whether and how the application of quantum analytical methods and ideas to a form of money showing dualistic features could be used to extract valuable economic information.
This article studies demand deposits (which represents the prevailing form of money in all contemporary economies) as a possible form of “money quanta", that is, quantities reflecting some aspects tipical of quantum physics, and analyzes it using a quantum model to describe some relevant aspects of it, including for instance how banks’ power to create money (i.e., issue deposits typically via the credit channel) is affected by the interactions taking place between the banks and between the banks and their environments, and how system efficiency and stability change in relation to changes in bank behaviors. This article lays the foundations of money quanta and is the first step towards a better understanding of the relevance of quantum mechanical tools and ideas in connection with the study of economics and finance. 
	\end{abstract}

	\vspace{2cm}
	
	
	\vspace{1cm}
	
	
	\vspace{2cm}
	
	
	\vfill


	\newpage
	
	\section{Introduction}
	
Demand deposits issued by commercial banks to their clients (henceforth, “bank deposits” or simply “deposits”) are considered and accounted for as debt liabilities of the issuing banks. They represent the counterparts to the value that depositors keep on the deposit accounts held with those banks. Thus, deposits are claims of the clients on their banks.  
Yet, how does the nature of debt obligations of bank deposits relate to the circumstance that a large share of such deposits are never going to be redeemed for cash or against settlement instruments under most circumstances (with the exception of extreme cases like runs on the banks)?
The Accounting View of Money (AVM) shows that, in fractional reserve regimes (see below), banks deposits take a hybrid form, whereby only a share of them can be regarded (and accounted for) as debt, while the remaining non-debt share represents net worth (equity) of the issuing banks, \cite{BC2021}. The AVM\footnote{The AVM challenges the long-established convention that fiat money is a debt obligation of the issuing entity. Debt involves obligations to transfer economic resources from borrowers to lenders, \cite{{iasb2018}}. Yet, no obligations derive today for the state to the public holding its coins, and no obligations derive for the central bank to the financial institutions holding its banknotes and reserves. Similarly, in the case of commercial bank deposits, the circumstance that a (large) share of such claims circulating in an economy are never going to be redeemed for cash or other settlement instruments (even under most critical circumstances) makes deposits a “hybrid” financial instrument, comprising both debt obligations and non-debt claims. All such non-debt money claims are a source of income for their issuers, which originates from the latter’s power to issue them at a cost that is lower than the attendant revenue.
	Challenges to the conventional view of state fiat money as debt are found in the work of BIS’s Archer and Moser-Boehm, \cite{archer}, who argue that banknotes “…act more like equity capital than debt obligations. As they bear no interest and are perpetual in character, they provide a stable funding base for income generation. To the extent that net income can be retained when needed, a large share of banknote liabilities provides a base for rebuilding equity if it has been depleted by a negative shock” (pp. 33–34).
	In line with the AVM, \cite{zell} observes that central bank money no longer obliges the central bank to exchange it for an asset other than central bank money and concludes that money created permanently (at the full discretion of the central bank) over the course of outright transactions can no longer be considered as debt and should be instead regarded as a form of equity.
	Furthermore, Bank of England’s Michael Kumhof and colleagues, \cite{kumhof}, concur with the AVM that central bank money cannot be characterized as a liability of the central bank since, in fiat money systems, the central bank is under no legal obligation to repay money holders in anything other than central bank money and consider it as a form of “social equity.”
	For the record, Covick and Davis, \cite{cov}, should be recognized as the first authors who claimed against considering banknotes as state liabilities, noting that they should be accounted for as a source of revenues (seigniorage), with implications for the public debt to be considered. And still for the record, it is Dyson and Hodgson, \cite{dyson}, who first argued that recording central bank money (banknotes and reserves) as a liability of the central bank is a throwback to the gold era of the 17th-19th centuries and showed that, according to current international accounting standards, state money should be recorded as equity of the state, not debt.
	The AVM moves the bar one level up and extends this conclusion to bank deposits, as said above, based on stochastic considerations, well supported by the recent pronouncement by the International Public Sector Accounting Standards Board, the international standards setting body for the public sector, \cite{ipsasb}. 
	The AVM and the similar views briefly recalled still have to make significant inroads within the scientific community before supplanting the anachronistic view of fiat money-as-debt. The purpose of studies like this on “money quanta” is to show what new insights can be gained by building on them.
} also shows the nature of deposit-issuing banks as hybrid institutions (part creators of money, and part pure financial intermediaries).

More broadly, the AVM resolves the apparent inconsistency between the formal rules of reporting money liabilities in the financial statements of the issuing banks and the economic substance of the money liabilities to be reported in the statements. Its main motivation lies on defining the correct representation of money in accounting terms and on drawing the implications that the correct representation of money would bring to bear for economic analysis – noting that "incorrect" are all representations of economic phenomena that are not consistent with the economic context as it has evolved over time, and therefore do no reflect the substance of the economic phenomena as they have changed with the evolution of the underlying context.

In this study, the focus is not on Accounting, but rather on using the important results that the application of modern Accounting to money makes possible, as the AVM does, to explore the association of the “hybrid” nature of bank deposits with the “dual” nature that characterizes the objects of quantum physics, and to investigate whether and how the application of quantum analytical methods to money with dualistic features can be used to extract valuable economic information. 

In other words, this study is about the exploration of bank deposits – the still largely prevailing form of money in all modern economies – as a possible form of money quanta\footnote{We decided to use this terminology in liew of the more obvious "quantum money", since the latter already exists and is defined as  as the quantum cryptographic protocol that is used to create and verify banknotes that are resistant to forgery.},  that can be analyzed through the methods of quantum mechanics. Its purpose is to see what informational benefits can be gained from the new approach. The study proposes a model, based on quantum mechanical ideas and rules, which aims to describe some relevant aspects of quantum money, and in particular some of its dynamical aspects.

The study is organized as follows. Section II briefly recalls the AVM, its analysis of the process of money creation by commercial banks, and its theoretical findings about the dual nature of bank deposits. Based on these findings, Section III investigates the analogies between bank deposits and quantum objects, and discusses the merits of analyzing bank money through quantum mechanics tools. The section also describes a simple quantum model for readers who are not familiar with quantum systems. In Section \ref{TMOD} we propose our quantum model of two banks interacting between them and with their own environments. The model will be based on operators of a certain nature (fermionic operators, \cite{rom}), which can be efficiently used to describe exchange between various agents\footnote{From this point of view, our quantum-like model has the advantage to describe naturally discrete exchanges, and to reflect, for their nature, the dual nature of money quanta. }.   Section \ref{sectSNS} contains some numerical simulations describing different situations: banks interacting with their environments, but not between them; banks mutually interacting, but not interacting with their environments and, finally, a {\em richer} situation in which all interactions are active. Section \ref{sectSNS} also contains our comments on the results, and our suggestions for future improvements of the model. Section VI concludes the article by emphasizing the innovativeness of the proposed approach and offers a perspective of future work in this area. Some aspects of our research are contained in two appendices.

\section{The Hybrid Nature of Bank Deposits}

\subsection{Foundations}

Deposits are claims on the issuing banks, which commit the latter to make available to depositors either cash or funds that are used to settle obligations to third-party agents. Banks create money in the economy by lending or selling deposits against securities or promises to, respectively, return or resell the deposits received (see Section \ref{box1}). 

When banks create money, they do not need to raise deposits in order to lend or sell deposits (Werner, 2014). Yet, they must avail themselves of the cash and reserves (held at the central bank) necessary to guarantee cash withdrawals from clients and to settle obligations emanating from client instructions to mobilize deposits in order to make payments or transfers, \cite{Bossone2020}.\footnote{Most contemporary payment settlement systems require that settlement takes place in central bank money. However, the principle of using safe assets for settlement is also adopted by those systems where central bank money is not available, and settlement can only happen in commercial bank money. Here, participating banks must first elect the money liability issued by one of them – typically, the one that is most highly reputed and financially solid and large – and then hold enough balances of such money for settlement purposes (CPSS, 2013). }   

The relevant payment orders are only those between clients of different banks, since the settlement of payments between clients of the same bank (i.e., "on us" payments) does not require the use of reserves and takes place simply by debiting and crediting accounts held on the books of the same bank. For cash withdrawals and interbank payments, every bank must determine the optimal amount of cash and reserves needed to cover deposits. These reserves consist of: i) cash reserves and reserves deposited with the central bank; ii) reserves from settlement of incoming payments from other banks; iii) borrowings from the interbank market; iv) borrowings from the central bank; v) immediate liquidation of unencumbered assets in the balance sheet, and vi) new deposits of cash from old and new clients. The new non-cash deposits from clients can only consist of deposits transferred from other banks, which fall under item ii) above. 

\subsection{Selling and Lending Deposits}\label{box1}

In a general sense, and for purposes of analogy with the sale of goods, deposits are "sold" whenever they are exchanged for other forms of value (e.g., goods, services, real or financial assets, or other currencies) and their ownership changes hands, either permanently or temporarily. The sale of deposits or of any other forms of value involves an "exchange transaction" as an operation where one party sells to (or purchases from) another party some value in exchange for some other form of value. The case of repurchase agreements, for instance, points to how a lending operation can be structured, both legally and economically, as a temporary sale of deposits with an obligation from the seller to repurchase the deposits at a future date. A "sale" of deposits is involved when they are issued against in exchange for funds, securities or credit claims.

Money leasing or lending transactions can be conceptualized as "sale" transactions, such as, for instance, when depositors lend funds to banks in exchange for deposits at (deposit claims on) the banks, or when banks lend deposit claims on themselves to borrowers. In these cases, too, the ownership of the funds loaned does change hands like in all sale/purchase transactions until transactions of opposite sign take place and offset the effects of the previous transactions – such as when funds depositor withdraw cash or request their banks to make payments from their accounts or when borrowers repay their debts. 

When a bank lends deposits, the operation can be broken into the following transactions: i) the bank sells new deposit claims by crediting the account of the borrower with the loan amount; ii) the borrower sells deposit claims back to the bank in exchange for funds, whenever she has to make payments to agents holding accounts with other banks; iii) the borrower resells the funds to the bank when her debt is due, in exchange for the extinction of her debt obligations.\footnote{For an application of the money “sale” concept to today’s digital currencies and the related legal underpinnings, see \cite{BC2021}.}

\subsection{Back to the hybrid nature}

Bank deposits constitute a debt liability for the issuing banks, since the latter are under an obligation to convert deposits into cash on demand from their clients or to settle payments in central bank reserves at the time required by payment system settlement rules.
 
However, in fractional reserve regimes, banks (by regulation or deliberate choice) only hold a fraction of reserves against their total deposit liabilities, and the amounts of reserves they actually use for settling interbank obligations are only a fraction of the total transactions settled\footnote{The expression “fractional reserve regimes” is here used to refer to all banking regulations that allow or require commercial banks to hold only a fraction (i.e., more than 0 \%  and less than 100\%) of the funds deposited with them by customers in the form of reserve balances held with the central bank. Positive reserve balances are necessary for banks to be able to settle interbank payments (as typically required by central banks) and to face meet liquidity outflows, \cite{gray}. Banks’ demand for reserve balances is positive even if no specific reserves ratio is legally imposed on them by the central bank. In jurisdictions with no such legal requirement (e.g., New Zealand, United Kingdom), and even in those systems where banks are required to hold zero reserves at the end of the operating day (e.g., Canada), one may still talk of "fractional" reserve regimes since the fraction of reserves voluntarily held by banks against their deposit liabilities is invariably greater than zero and less than 100\%. Notice that standards for banks’ assets and liabilities management recommended under the Basel framework are aimed for banks operating in fractional reserve regimes, which nowadays underpin most (if not all) commercial banking legislations worldwide. Even when subjecting banks to the “net stable funding ratio,” defined as the amount of available stable funding relative to the amount of required stable funding, the latter (which is typically a larger aggregate than reserve balances) is only a portion of their total assets. Besides, for the purpose of allowing banks to economize on costly reserves (where these are not enough remunerated), the standards provide for banks to hold other high-quality liquidity asset that earn some interest to them. In general, the rationale for the Basel standards rests on the recognized need to limit the financial fragility inherent in all banks operating in fractional regimes – no such standards would be required for narrow banks or if banks were not involved in liquidity and maturity transformation. Even when these standards provide for tighter requirements for large (or systemically important) banks, this does not contradict the fact that (all else being equal) larger banks benefit from economies of scale in the use of (costly) reserves to support their operations.  }. The more limited is the use of cash in the economy, and the larger the economies of scale (EoS) in the use of reserves (as permitted by payment system rules and clients’ non-simultaneous mobilization of deposits and cash withdrawals), the lower is the volume of reserves that banks need to back up the issuance of new deposits. 

Payment system rules affect the use of reserves via two channels: the settlement modality (that is, netting or gross settlement) and the technology adopted. Modern technologies (re)introduce elements of netting into gross settlement processes and increase the velocity of circulation of reserves, thereby allowing banks to economize on the use of reserves for any given volume and value of payments settled.  In the hypothetical case of a fully consolidated banking system in a cashless economy, where all agent accounts sit with only one bank, all payments and transfers would be "on us" for the bank. The bank would therefore need no reserves for settling transactions and would be under no debt obligation to its clients, and it might create all the money that the economy would be willing to absorb, without any need for holding reserves. In such case, the money would have the same power as legal money in settling all debts.  

In real-world economies, however, there are multiple banks whose payment activities generate interbank settlement obligations. Yet, the fractional reserve regime and the EoS that are made possible both by payment system rules and by depositors’ asynchronous mobilization of deposits and cash withdrawals reduce the volume of reserves needed by the banks to back their liabilities (debts). Under increasing EoS, banks can create more liabilities (by lending or selling deposits) with decreasing reserve margins for coverage. From both the hypothetical case above and this last observation follows that, all else equal, a more consolidated banking system affords individual banks lower coverage of their liabilities (and at lower cost) than a less concentrated system would. 
 
More generally, absent (very extreme) adverse economic or market contingencies inducing depositors either to convert their deposit claims into cash or to transfer them across banks, the liabilities represented by deposits only partly constitute debt liabilities of the issuing bank, which as such require cash and reserve coverage. 

By way of example, assume that interbank liabilities are settled with reserves and banks create and issue deposit claims by lending them to clients at an interest. Under a fractional reserve regime, the outstanding stock of the deposits created by the banks is backed only partially (fractionally) with (costly) reserves (ever under most market stress circumstances). In this case, the “unbacked” stock of deposits (i.e., the stock of interest-earning loaned deposits net of the stock of costly reserves for coverage), which banks can create at will (until they find borrowers willing to borrow at the given terms and conditions), is a free resource that banks have used to “buy” revenue earning assets (i.e., the interest earning credit claims). In the banks’ balance sheet, these real claims are recorded on the assets side, until they are repaid) and the free resource created (as determined above) should be recorded as equity.

In other words, in fractional reserve regimes, the uncovered part of the banks’ deposit liabilities are a free resource that did not exist before its creation. Once created, it can be used to buy claims on real resources (say, loans generating capital and interest payments). This free resource is a source of income accruing to the deposit issuing banks. To the extent that this income is accumulated and undistributed, it is equivalent to equity and should be accounted accordingly. Deposits, therefore, consist of "debt-deposits" and "equity-deposits." This (admittedly controversial) proposition is further explained in \cite{BC2021}, especially Boxes 1-3.  

This hybrid (dual) nature of deposits is stochastic in as much as, at the time of issuance, every deposit unit can be either a debt-deposit (if, with a certain probability the issuing bank receives requests for cash conversion or interbank settlement) or an equity-deposit (with complementary probability). Faced with such stochastic double nature of its money, a commercial bank finds it convenient to provision the deposit unit issued with a level of reserves that equals only the expected value of the associated debt event (possibly augmented by some unexpected variation margin), rather than the full value of the deposit unit issued - indeed, the origin and rationale of fractional reserve regimes in banking.

Here, "stochastic" refers to the fact that – ex ante – a bank creating one unit of deposit expects (probabilistically) that only some share of that unit will translate into debt, while the remaining share (still probabilistically) will not be subject to requests for conversion into cash or reserves. The share of debt-deposits (or equity-deposit, as its complement) is a stochastic variable that is influenced by behavioral and institutional factors (for example, cash usage habits and payment system settlement rules) as well as by contingent events. In times of market stress, the share of debt-deposits tends to increase, while it tends to be low when trust in the economy (and the banking system in particular) is strong. Policy and structural factors that strengthen such trust (for example, the elasticity with which the central bank provides liquidity to the system when needed, or a deposit insurance mechanism) increase the share of equity-deposits.

This argument is evident when applied to the whole banking system, but it holds also for each individual bank, albeit to different extents depending on the size of each bank (where size refers to the volume and value of payment transactions that the bank intermediates relative to the total payment transactions in the system).  From the discussion so far it follows that, all else being equal, the stochastic share of debt-deposits for small banks is greater than for larger banks. Vice versa, the larger is the bank, the greater is the share of equity contained in its deposit liabilities.  
In concluding, while the conventional view holds that bank deposits are a liability (debt) of the issuing banks, the AVM argues that, in a fractional reserve regime, they are only partly debt obligations of the issuing bank, with the residual part constituting equity of the same bank in the form of accumulated and undistributed income.\footnote{Such equity, however, corresponds to the accumulated and undistributed income generated by deposits and must not be confused with bank’s capital (with its attendant rights and obligations).}

\section{Bank Deposits as money quanta}

Following from the preceding considerations, while conventional views of money hold that bank deposits are a liability (debt) of the issuing banks, according to the AVM, in a fractional reserve regime only a part of them represents debt obligations of the issuing bank, while the other part constitutes equity of the issuing bank in the form of accumulated and undistributed income.\footnote{Such equity, however, corresponds to the accumulated and undistributed income generated by deposits and must not be confused with bank’s capital (with its attendant rights and obligations).} In other words, deposits are in all cases claims on the issuing bank, which can take different forms depending on whether they stand (stochastically) to be converted into cash or to be used for payments (and require central bank reserves for settlement). If they do, they are liabilities of the issuing bank;\footnote{It should be noted that while economists typically consider the balance sheet as consisting of “assets” and “liabilities”, accountants correctly consider the balance sheet as consisting of “resources” and “claims”, with the latter including “liabilities” (i.e., debts) and “equity”.} if they don’t, they are equity of the issuing bank. Ex ante, they would be both and the two forms would overlap (in a kind of superposition of states in quantum physics), until some external events caused them to collapse in one form or the other\footnote{Notice that, in the language of quantum foundations, the debt vs. equity distinction is ontic, that is, it has a real existence since, ex ante, deposit claims held in bank accounts stand a probability to be redeemed into something else (cash or central bank reserve to be transferred), thereby implying a debt obligation for the issuing banks, and they stand a complementary probability of never being redeemed into anything else , and therefore they are not liabilities of the issuing banks, \cite{ipsasb}. The collapse into one form or the other is determined exogenously to the issuing banks. True, as discussed below, different behaviors of the issuing banks may alter the debt/equity compositions of their deposit claims; yet,  the collapse of such claims into one of their two forms depends only on claim holders' choices of, say, redeeming their deposit claims into cash or using them for payments. The latter may require their banks to  use  reserves for settling the related interbank obligations.   }. This looks very much the same to what happens in quantum mechanics, when a given system is in a state that is (usually) a superposition of other states (spin up and spin down, first or second energetic level, and so on), but when some measurement is performed on the system, which leads to its collapse in a specific state that corresponds to the result of the measurement: this is an "eigenstate" of the quantity of the system (an Hermitian operator, in mathematical terms) that is being measured. 

The hybrid nature of bank deposits may be subjected to different dynamics, depending on the characteristics of the issuing banks, their interactions with the environment where they operate, and their mutual interactions within the payments system. 

The role of bank size was discussed already, with large banks enjoying a larger potential for issuing equity-deposits and smaller banks being more constrained to rely on debt-deposits. Yet, all else equal, the debt/equity mix of the deposits issued by the same banks can change as a result of changes in, say, the risk attitude of banks. A bank (of any size) showing greater risk tolerance, and therefore wanting to take a more aggressive stance on lending, would issue more deposits for any given volume of its cash and central bank reserve holdings, thus raising the relative share of equity-deposits over the total deposits issued, and vice versa for lower risk tolerance and a more conservative lending policy. This is important from the financial stability standpoint: whereas high money creation power can be the sign of a soundly managed bank that achieves efficiency in use of reserve and is exposed to low deposit liabilities, the same power may be abused if the bank behaves unsoundly, exposing itself to poor asset quality and large (uncovered) liabilities.   

Similar effects would follow from changes in the public sentiment characterizing the economic environment where the banks operate. The diffusion of positive news and related rise of optimism within a banking business community might induce banks (of all sizes) to extend more loans for given cash and reserve levels, again, raising their relative share of equity-deposits over their total deposits.\footnote{In all the above examples, nothing ensures that the ex-ante (expected) mix of debt/equity deposits would coincide with its ex-post (actual) value, since the payments activity of banks may require them to demand more reserves for settlement purposes than they had initially planned.} An important additional effect, in this case, would derive from the (positive or negative) spillovers generated by the banks’ strategic interactions within the payments system: if, under optimism, each bank expects the others to expand their lending, it will figure that reserves will circulate more rapidly in the payments system, with more reserves accruing to them as they will receive more payments. This means that more deposits can be issued for a given volume of reserves circulating in the payments system, thus raising the share of equity-deposits over total deposits. Obviously, the reverse would happen under pessimistic expectations and negative spillovers from bank interactions.   
Finally, the payments system, too, may have an impact on the debt/equity deposit mix. As an example, the introduction of liquidity-saving automated solutions re introduced in real-tien gross settlement systems, which allow banks to economize on central bank reserves for given volumes and values of payment transactions to be settled, would tend to raise the (optimal) equity-deposits ratio of banks (all else being equal).

Having shown the hybrid (dual) nature of bank deposits and their stochastic underpinning, as well as the superposition of states they embody and the possible dynamic evolution of their associated states, it is more than tempting to think of deposits as quantum-like objects – or as a form of “quantum money” – which would therefore be susceptible of being treated analytically with the mathematical apparatus of quantum mechanics. What advantages would such an approach bring to bear for economic analysis, such as for instance for financial stability and policy purposes?\footnote{This exercise would still be worth doing even if only under an “as if” assumption. That is, considering that the AVM idea of bank deposits as hybrid financial instruments (partly debt, partly equity) may still encounter resistance (despite the AVM being grounded on extra solid principles, with scholars and experts moving in its direction), it would still be useful for the “resistants” to suspend their judgment and accept the quantum exercise assuming for a moment that the AVM holds true, and see what the results would be.}

For readers of this article who are not familiar with quantum physics, it might be useful to know that at the beginning of the twentieth century, scientists were unveiling one of the most well-kept natural mysteries – the dual wave particles or the dual nature of matter and energy. Initially, and for a long time, the properties of a material object or light were defined by their respective particle or wave structure. It was then discovered by experiments that matter has wave characteristics, much as light shows behaviors that can only be associated with a particle structure. Therefore, both were said to have a “dual” nature, that is, a nature that features both particle and wavelength characteristics. 
The concept of wave-particle duality of matter and energy is the essence of quantum mechanics, the branch of Physics that studies microscopic objects. Quantum mechanics refers to the physical properties of matter at the scale of an atom or subatomic particles. It differs from classical physics in that energy, momentum, and other quantities of a system are usually restricted to discrete values (quantization); objects, as said, have characteristics of both particles and waves; and there are limits to how accurately the value of two physical (conjugate) quantities, like the position and the momentum of a given particle, can be efficiently measured, given a complete set of initial conditions (the so-called “uncertainty principle”).
In fact, a fundamental feature of the quantum mechanics is that it usually cannot predict with certainty what will happen, but only assign probabilities to different events. For example, a quantum particle like an electron can be described by a wave function, which associates to each point in space a probability amplitude to find the electron in a given position in space.  
One consequence of the mathematical rules of quantum mechanics is a tradeoff in predictability between different measurable (again, conjugate) quantities. The most well-known version of the uncertainty principle says that no matter how a quantum system is prepared, or how carefully experiments upon it are arranged, it is impossible to have a precise prediction for the measurement of its position and, at the same time, for the measurement of its momentum. This is a consequence of the properties of the position $\hat x$ and momentum $\hat p$ not to commute: $\hat x\hat p\neq \hat p\hat x$. This is necessary since, to explain experimental data, quantum variables are time-dependent {\em operators} rather than {\em functions}. A consequence of this non commutativity of the observables of quantum systems is that, the more precise is our knowledge of the position of the quantum particle, the less we can say about its momentum, and vice versa. This particular feature, which is intrinsic with quantum mechanics, is easily met in other, and completely different, contexts. For instance, decision making under uncertainty is one of such a contexts: we have to decide on some specific problem, but we have some uncertainty (like noise, rumours, external inputs,...) which make it difficult to come out with a clear and unique decision. This is just one of the many appearances of some quantum-like feature outside quantum mechanics. Many authors, in different fields, are now interested in using quantum tools in these, apparently non standard, realms. Busemeyer, Khrennikov, Pothos, Haven, Baaquie, \cite{Busemeyer2012,baa,havkhrebook,Khrennikov2010} are just some of them.
Bagarello (2012 and 2019) illustrates how quantum mechanics can indeed be applied to classical problems in finance, biology, economics, and elsewhere, and how quantum tools – in particular, the number operator $\hat N$ and the ladder operators giving rise to $\hat N$\footnote{In quantum mechanics, for systems where the total number of particles may not be preserved, $\hat N$ is the observable that counts the number of particles, which can be created by a {\em creation operator} $c^\dagger$, or annihilated by an {\em annihilation operator} $c$, and $\hat N=c^\dagger c$.} – can be used to create dynamical systems, like the one represented in the model discussed in this article, where the variables are operator-valued functions. The original driving idea was that the lowering and raising operators related can be used in the description of processes where some relevant quantities change discontinuously. To cite a few examples, stock markets, migration processes or some biological systems: multiples of share are exchanged in a market; one, two or more people move from one place to another; and one cell duplicates producing two cells. This suggests that objects labeled by natural numbers are important, in some situations. 
More technically, nonnegative integers can be seen as eigenvalues of some suitable number operator constructed using ladder operators, and ladder operators can be efficiently used to describe systems where discrete quantities of some kind are exchanged between different agents. Hence ladder operators, and some combinations of them, can be used in the description of specific systems, and this is indeed the core of what we will see in the rest of this article. 

The advantages of treating bank deposits as quantum objects are, at least twofold. First, whereas under the traditional view the claims side of a bank's balance sheet feature a clear distinction between liabilities and equity, with demand deposits sitting squarely on the former item line, the hybrid nature of deposits uncovers a different and richer underlying economic reality that, as the AVM shows, bears new accounting and financial implications, including for the different forms of income that bank's money creation power generates and should be appropriately recognized, \cite{BC2021}.  Second, the hybrid nature of deposits and their tractability as quantum objects allows to incorporate within their analysis a broader range of risk and
uncertainty dynamics that might uncover elements that would not be evident under traditional analysis. In the following sections, this study – as the first of its genre – only seeks to suggest that this might indeed be a new field of theoretical and empirical investigation, inviting for subsequent contribution, while certainly not claiming to offer an exhaustive analysis of all possibilities.

To clarify what we have written so far, we devote the following section to a simple predator-prey model, which can be skipped by all readers who are familiar with quantum systems and fermionic operators.

\subsection{A simple predator-prey model}\label{sectpredprey}

Let us consider a system $\mathcal{S}$, having two (fermionic) degrees of freedom, describing two different agents of $\Sc$, and described by the Hamiltonian
\begin{equation}
	H=H_0+\lambda H_I,\qquad H_0=\omega_1a_1^\dagger a_1+\omega_2a_2^\dagger a_2, \quad H_I=a_1^\dagger a_2+a_2^\dagger a_1,
	\label{MM23}
\end{equation}
where $\omega_j$ and $\lambda$ are real (and positive) quantities in order to ensure that $H$ is Hermitian. This is a sufficient condition to guarantee that its eigenvalues, which are the quantities measured in a concrete experiment, are real. The operators $a_j$ and $a_j^\dagger$ are assumed to satisfy the following canonical anti-commutation relation (CAR)\footnote{It might be interesting to remind that $a_j^\dagger$ is just the complex conjugate and transposed version of $a_j$: this is the effect of the {\em adjoint} operation $\dagger$, here. More complicated would be its effect of operators acting on infinite dimensional Hilbert spaces, \cite{Bagarello2012,Bagarello2019} and references therein.}:
\begin{equation}
	\{a_i,a_j^\dagger\}=\delta_{i,j}\,\1,\qquad \{a_i,a_j\}=\{a_i^\dagger,a_j^\dagger\}=0,
	\label{MM21}
\end{equation}
$i,j=1,2$, where, as usual, $\1$ is the identity operator, and where $\{x,y\}=xy+yx$ is the anti-commutator between $x$ and $y$. 
Of course, when $\lambda=0$, the two agents of $\Sc$ are not interacting. Because of their properties, see \cite{Bagarello2019} for instance, the various terms in $H$ have the following interpretation: the contribution $a_1^\dagger a_2$ in $H_I$ describes the fact that the density of species 1 is increasing (since $a_1^\dagger$ is a raising operator), while the density of species 2 is simultaneously decreasing (since $a_2$ is a lowering operator): hence $a_1^\dagger a_2$ can be understood as an increasing of predators which causes a decrease of preys. The other term in $H_I$,  $a_2^\dagger a_1$, describe the opposite phenomenon: when the predators decreases (because of $a_1$, which is a lowering operator), the preys may increase (and this is reflected by the presence of $a_2^\dagger$, which is a raising operator). If $\lambda=0$ both these mechanisms are absent, and $H=H_0$. But $H_0$ only contains pairs of raising and lowering operators of the same species. So $H_0$ is only responsible of {\em counting} the members of each species: the densities of the two species do not change. 

Indeed, this a priori interpretation is exactly recovered after our explicit computations. Let us briefly show how, starting with the definition of the Hilbert space (i.e., the particular vector space where the model is defined) we work with.

The operators $a_i$ and $a_i^\dagger$ give rise, in this case, to a simple four-dimensional Hilbert space, \cite{rom}. First, the eigenstates of the number operators $\hat n_j:=a_j^\dagger a_j$ are easily obtained: if $\varphi_{0,0}$ is the {\em ground vector} of $\mathcal{S}$,
$a_1\varphi_{0,0}=a_2\varphi_{0,0}=0$, an \index{Basis!orthonormal}o.n. basis of the four-dimensional Hilbert space $\mathcal{H}$
of $\Sc$ is given by the following vectors:
\begin{equation}
	\varphi_{0,0},\qquad \varphi_{1,0}:=a_1^\dagger\varphi_{0,0}, \qquad \varphi_{0,1}:=a_2^\dagger\varphi_{0,0}, \qquad \varphi_{1,1}:=a_1^\dagger a_2^\dagger\varphi_{0,0}.
\label{add1}\end{equation}
We have
\begin{equation}
	\hat n_1\varphi_{n_1,n_2}=n_1\varphi_{n_1,n_2},\qquad \hat
	n_2\varphi_{n_1,n_2}=n_2\varphi_{n_1,n_2}. \label{MM22}
\end{equation}

The equations of motion for the annihilation operators $a_j(t)$ are deduced using the Heisenberg approach: if $X(t)$ is an operator-valued unknown of the system $\Sc$, associated to an Hamiltonian $H$, then $\dot X(t)=i[H,X(t)]$ is the differential equation for $X(t)$, which should be solved to deduce the time dependence of the variable $X$. Notice that we have $[x,y]=xy-yx$, the commutator between $x$ and $y$. With this in mind, we deduce 
\begin{equation}
	\dot a_1(t)=-i\omega_1 a_1(t)-i\lambda a_2(t),\qquad
	\dot a_2(t)=-i\omega_2 a_2(t)-i\lambda a_1(t),
	\label{MM24}
\end{equation}
which  can be solved imposing the \index{Initial conditions}initial conditions $a_1(0)=a_1$ and $a_2(0)=a_2$, and the solution is
\begin{equation}
	\begin{aligned}
		&a_1(t)=\frac{1}{2\delta}\left(a_1\left((\omega_1-\omega_2)\Phi_-(t)+\delta\Phi_+(t)\right)+2\lambda a_2\Phi_-(t)\right),\\
		&a_2(t)=\frac{1}{2\delta}\left(a_2\left(-(\omega_1-\omega_2)\Phi_-(t)+\delta\Phi_+(t)\right)+2\lambda a_1\Phi_-(t)\right),
	\end{aligned}
	\label{MM25}
\end{equation}
where
\begin{equation}
	\begin{aligned}
		&\delta=\sqrt{(\omega_1-\omega_2)^2+4\lambda^2},\\
		&\Phi_+(t)=2\exp\left(-\frac{it(\omega_1+\omega_2)}{2}\right)\cos\left(\frac{\delta t}{2}\right),\\
		&\Phi_-(t)=-2i\exp\left(-\frac{it(\omega_1+\omega_2)}{2}\right)\sin\left(\frac{\delta t}{2}\right).
	\end{aligned}
\end{equation}
Then, the functions $n_j(t):=\left<\varphi_{n_1,n_2},
\hat n_j(t)\varphi_{n_1,n_2}\right>$ are
\begin{equation}
	\label{MM26-27}
	\begin{aligned}
		&n_1(t)=\frac{n_1(\omega_1-\omega_2)^2}{\delta^2}+ \frac{4\lambda^2}{\delta^2}
		\left(n_1\cos^2\left(\frac{\delta t}{2}\right)+n_2\sin^2\left(\frac{\delta t}{2}\right)\right), \\
		&n_2(t)=\frac{n_2(\omega_1-\omega_2)^2}{\delta^2}+ \frac{4\lambda^2}{\delta^2}
		\left(n_2\cos^2\left(\frac{\delta t}{2}\right)+n_1\sin^2\left(\frac{\delta t}{2}\right)\right),
	\end{aligned}
\end{equation}
which oscillate in time.

These functions could be interpreted, in agreement with other similar applications, as the densities of two species, $\mathcal{S}_1$ and $\mathcal{S}_2$, interacting as in (\ref{MM23}) in a given (small) region\footnote{Other interpretations are also possible.
	For instance, they can play the role of the {\em decision functions} of two interacting agents, trying to decide on some binary
	question.}. The interaction Hamiltonian $H_I$ in (\ref{MM23}) describes, as already stressed, a sort of \index{Predator-prey!mechanism}predator-prey mechanism, because of the presence of a creation operator for one species coupled (i.e., multiplied) to an annihilation operator for the other species.  This is reflected by the solutions in (\ref{MM26-27}), which show how the two densities, because of the interaction between  $\mathcal{S}_1$ and $\mathcal{S}_2$, oscillate in the interval $[0,1]$. Notice that, however, if $\lambda=0$, $n_j(t)=n_j$: the densities stay constant, and nothing interesting happens in $\mathcal{S}$. We refer to \cite{Bagarello2012} for more details on this simple model. Here we only want to stress that the same idea briefly described here is the core of what will be done in the rest of the article, in our financial context.

\vspace{3mm}

It is therefore only natural to think of applying these tools to the creation or destruction of bank deposits – as quantum objects – trying to predict dynamics relating to what part of them is debt and becomes equity, or vice versa, due also to the interactions among the issuing banks and between the banks and their surrounding environment.

\section{The model}\label{TMOD}

Before describing the model, readers should be informed that Appendix A lists the main variables and parameters of the model (both their denomination and formal - symbolic - representations), explains their meaning in Physics, and interprets them in terms of Economics. 
The system consists of two banks $\B_1$ and $\B_2$, and two related sets of information, each reflecting the "environment" where each bank operates, comprising factors that are external to the bank and influence its operations, such as, inter alia, the systemwide and local (or sectoral) macro-financial conditions underpinning the bank’s activity, market expectations, bank reputation with the public, the size of the market where the bank operates relative to the whole economy, the supervisory and regulatory regime under which the bank operates, and the payment system rules that shape the its money creation power, as discussed in Section 1.2. We call these two sets $\Bc_1$ and $\Bc_2$. What we describe in this model is the ratio between the {\em debt-deposits}, $DD$, and the {\em total deposits issued by the bank} $D$: $r^D=\frac{DD}{D}$, where $D=DD+DE$ and $DE$ is {\em equity-deposits}, Hence $r^D\in[0,1]$. In particular, $r^D\simeq0$ for very large banks and $r^D\simeq1$ for smaller banks.

Both scenarios are deliberately extreme and for analytical purposes only. They resemble two extreme cases, and reality would always fall somewhere in between. The case of $r^D\simeq0$  would be one of a fully consolidated banking system with only one bank operating in a cashless economy, where all payments would take place be on the books of the same bank (“on us” payments). In this case, the bank could in principle operate with zero reserves since no reserves would be necessary to settle payments between its customers. Indeed, the central bank could still impose on the bank a reserve or other liquidity or capital requirements, for monetary or credit control purposes, but this would not contradict the EoS that the system would make possible to the bank. This case, thus, exemplifies the case of a very concentrated banking (and payment) system. At the other extreme, $r^D\simeq1$ , would be one of an atomistic banking sector (with hypothetically zero-dimensional banks), where every single bank could still issue new deposits as loans to borrowers or sales to, say, securities dealer, but where they would have to cover all new deposit issuances with central bank reserves. In other words, this would be the case where the fractional reserves regime would de facto be equivalent to a “narrow banking,” regime where the fraction is $\geq  100\%$ and every deposit unit must be backed by (at least) a unit of central bank reserves.

The reason why we consider here two banks is to allow for the possibility of describing banks with different $r^D$ at the same time. 
Each bank could be, at $t=0$, in two possible states, 0 and 1. The intermediate states are simply linear combinations of these. Hence,
we have four different possibilities, to which, following Bagarello \cite%
{Bagarello2015a,Baghavkhr}, we associate four different and mutually orthogonal
vectors in a four dimensional Hilbert space $\mathcal{H}_{\B}$.
These vectors are $\varphi _{0,0}$, $\varphi _{1,0}$, $\varphi _{0,1}$ and $\varphi _{1,1}$.  

The first vector, $\varphi _{0,0}$, describes the fact
that, at $t=0$, the two banks have both $r^D=0$: $r^D_1=r^D_2=0$. Of
course, such a choice can change during the time evolution of the system.
Analogously, $\varphi _{0,1}$ describes the fact that, at $t=0$, the first
bank has a $r^D_1=0$, while the second has $r^D_2=1$. And so
on. $\mathcal{F}_{\varphi }=\{\varphi _{k,l},\,k,l=0,1\}$ is an orthonormal
basis for $\mathcal{H}_{\B}$. The general vector of
the system $\B=\B_1\cup\B_2$ (i.e. of the two banks), for $t=0$,
is a linear combination
\begin{equation}
\Psi _{0}=\sum_{k,l=0}^{1}\alpha _{k,l}\varphi _{k,l},  \label{20}
\end{equation}%
where it is natural to assume that $\sum_{k,l=0}^{1}|\alpha _{k,l}|^{2}=1$
in order to normalize the total probability. Indeed, for instance, we
interpret $|\alpha _{0,0}|^{2}$ as the probability that $\mathcal{S}_{%
\B}$ is, at $t=0$, in a state $\varphi _{0,0}$, i.e. that both $%
\B_1$ and $\B_2$ have $r_D\simeq 0$. Analogous
interpretations can be given to the other coefficients.

We construct the states $\varphi_{k,l}$ by using two fermionic operators,
i.e. two operators $b_1$ and $b_2$, satisfying the following \textit{%
canonical anti-commutation rules} (CAR):
\begin{equation}
\{b_k,b_l^\dagger\}=\delta_{k,l}\,1 \!\! 1,\qquad \{b_k,b_l\}=0,  \label{21}
\end{equation}
where $k,l=0,1$. Here $1 \!\! 1$ is the identity operator and $\{x,y\}=xy+yx$
is the anticommutator between $x$ and $y$. Then we take $\varphi_{0,0}$ as
the vacuum of $b_1$ and $b_2$: $b_1\varphi_{0,0}=b_2\varphi_{0,0}=0$, and, as in (\ref{add1}),
we construct the other vectors out of it:
\begin{equation*}
\varphi_{1,0}=b_1^\dagger\varphi_{0,0}, \quad
\varphi_{0,1}=b_2^\dagger\varphi_{0,0}, \quad
\varphi_{1,1}=b_1^\dagger\,b_2^\dagger\varphi_{0,0}.
\end{equation*}

The operators in (\ref{21}) are useful since they
are used, together with other operators we will introduce later, to write down a Hamiltonian for the full system $\mathcal{S%
}=\B\cup\Bc$, where $\Bc=\Bc_1\cup\Bc_2$, from which we deduce the dynamics of the \emph{observables} of $\mathcal{%
S}$, i.e., the variables needed to describe $\mathcal{S}$, and $\mathcal{S}_{%
\B}$ in particular, via the Heisenberg rule. What we will do next extends what was done in Section \ref{sectpredprey} in a much simpler situation, but using similar ideas.

Let  $\hat n_{j}=b_{j}^{\dagger }b_{j}$ be the number operator of the $j$-th bank, which represents bank j’s net money creation power as determined by the combined effects of the creation and annihilation operators (defined above) (see Appendix A). 
Note that the CAR above implies that $\hat{n}_{1}\varphi _{k,l}=k\varphi
_{k,l}$ and $\hat{n}_{2}\varphi _{k,l}=l\varphi _{k,l}$, $k,l=0,1$. Then the eigenvalues of these operators
correspond to the value of $r^D_j$ at $t=0$. 

{Thus, $\hat n_{j}$ is the operator which, using the operator $H$ in (\ref{24}), describes, at an operational level, the time-evolved version of $r^D_j$, whose optimal value is determined by the environment where bank j operates (as discussed below). Counterintuitively, if $\hat n_{j}$ increases (i.e., $r^D_j$ increases), then the money creation power of bank j declines, and vice versa.} 

To go from operators to functions, we will further need to compute some suitable expectation values, as described below, see (\ref{230}). 

Our main effort now consists of \emph{giving dynamics} to the number
operators $\hat{n}_{j}$, for all banks, following the scheme described in \cite
{Bagarello2012,Bagarello2019}, and briefly sketched in Section \ref{sectpredprey}. Therefore, what we first need is to introduce a Hamiltonian
$H$ for the system. Then, we will use this Hamiltonian to deduce the
dynamics of the number operators as $\hat{n}_{j}(t):=e^{iHt}\hat{n}%
_{j}e^{-iHt}$, and finally we will compute the mean values of these
operators on some suitable state which describes the status of
the system at $t=0$ (see below). We refer to \cite{Bagarello2015a}-\cite{FFF} for the
details of our construction. Here we just recall that $H$ is the Hamiltonian
of an open system, since the two banks $\B_{1}$ and $\B_{2}$ interact in some way with their
environments $\mathcal{B}_{1}$ and $\mathcal{B}_{2}$, whose operators live in
an infinite-dimensional Hilbert space. This is because, in this way, we can use $\mathcal{B}_{1}$ and $\mathcal{B}_{2}$ to model many effects driving, or connected to, the effective value of $r^D_j$.

The full Hamiltonian of $\Sc$ we consider here, $H$, is the following:

\begin{equation}
\left\{
\begin{array}{ll}
H=H_0+H_I+H_{int}, &  \\
H_{0}=\sum_{j=1}^{2}\omega _{j}b_j^\dagger b_j+\sum_{j=1}^{2}\int_{\mathbb{R}%
}\Omega_j(k)B_j^\dagger(k)B_j(k)\,dk, &  \\
H_{I}=\sum_{j=1}^{2}\lambda_j\int_{\mathbb{R}}\left(b_j
B_j^\dagger(k)+B_j(k)b_j^\dagger\right)\,dk &  \\
H_{int}=\mu_{acm}\left(b_1^\dagger b_2+b_2^\dagger
b_1\right)+\mu_{cm}\left(b_1^\dagger b_2^\dagger+b_2 b_1\right). &
\end{array}%
\right.  \label{24}
\end{equation}

{Here $\omega _{j}$, $\lambda _{j}$, $\mu_{acm}$ and $\mu_{cm}$  are real 
quantities, and $\Omega _{j}(k)$ are real functions, $k\in\mathbb{R}$. In economic terms, $\omega _{j}$ measures the (marginal) cost that bank j sustains to change its money creation. The value of $\omega_j$ is lower, the larger the bank due to EoS,  and vice versa\footnote{A physical interpretation of this feature is the following: if we have a quantum system driven by an Hamiltonian $h=\omega b^\dagger b$, if $\omega$ is {\em small}, then it is easier to move from one energetic level to another, since $\omega$ is exactly the difference between the energy of the levels. On the other hand, if $\omega$ is large, then a larger amount of energy is needed to move from the ground to the excited level, and this transition is less probable.}; $\lambda _{j}$ measures the strength of the interaction between bank j and its own environment;  $\mu_{cm}$ measures the tendency of the two banks’ money creation powers to co-move (that is, to increase or decrease together); $\mu_{acm}$ measures the tendency of the banks’ money creation powers to move in opposite directions; and $\Omega _{j}(k)$ measures the relative importance of bank j’s environment relative to the system where the bank operates.} In analogy with the $%
b_{j}$'s, we use fermionic operators $B_{j}(k)$ and $B_{j}^{\dagger }(k)$ to
describe the environment:
\begin{equation}
\{B_{i}(k),B_{l}(q)^{\dagger }\}=\delta _{i,l}\delta (k-q)\,1\!\!1,\qquad
\{B_{i}(k),B_{l}(q)\}=0,  \label{23}
\end{equation}%
$k,q\in\mathbb{R}$, which have to be added to those in (\ref{21}). Moreover each $b_{j}^{\sharp }
$ anti-commutes with each $B_{j}^{\sharp }(k)$: $\{b_{j}^{\sharp
},B_{l}^{\sharp }(k)\}=0$ for all $j$, $l$ and $k$. Here $X^{\sharp }$
stands for $X$ or $X^{\dagger }$. A similar operator has been considered for other systems, and in particular in  \cite{Bagarello2015a,Baghavkhr}, with a different meaning but with a similar strategy.

The various terms of $H$ describes the following effects: i) $H_{0}$ is the \emph{free} Hamiltonian, which produces no
time evolution for the number-like operators $\hat{n}_{j}$,\footnote{In (\ref{230}) their time evoluted mean values will produce our main object of study, the so-called Debit functions.} since $%
[H_{0},\hat{n}_{j}]=0$; ii) $%
H_{I}$ describes the interaction between each bank and its own environment. The idea is that this interaction is one of the essential mechanism driving the value of $r^D_j$; and iii) $H_{int}$ describes two different interactions
between $\B_{1}$ and $\B_{2}$ . In particular, if we take $\mu_{cm}=0$, then the surviving term in $H_{int}$ describes an opposite behavior of the two banks: if the value of $r_2^\D$ decreases, because of $b_2$, that of $r_1^\D$ increases, due to the action of $b_1^\dagger$. This is the effect of $b_1^\dagger b_2$. The second contribution in $H_{int}$ (when $\mu_{cm}=0$), $b_1^\dagger b_2$, describes the opposite behavior: $r_2^\D$ increases while $r_1^D$ decreases. If we rather fix $\mu_{acm}=0$, what survives in $H_{int}$ describes a co-moving effect: both $r_1^\D$ and $r_2^\D$ either decrease or increase, simultaneously. Of course, these two choices of $\mu$'s are extreme ones. Yet, they determine the range within which our dynamic system may evolve over time, as the banks' environment changes and the banks themselves may behave differently in the market. Thus, it might easily happen that the values of $\mu_{acm}$ and $\mu_{cm}$ change during the time evolution of the system according, for instance, to some {\em rule} in the sense of $(H,\rho)$-induced dynamics, \cite{bagrules}. An analysis of these changes is postponed to future work. Here, we only consider the choice of constant $\mu_{cm}$ and $\mu_{acm}$, of which not necessarily one is zero: this, too, can be understood in terms of dynamics, since the operators $b_j$ and $b_j^\dagger$ are also (intrinsically) time dependent.

When $\mu_{cm}=0$, that is, when there is no tendency for the banks’ money creation powers to co-move (i.e., increase or decrease together), the
two $r^D$ change differently, while they change in the same way when $\mu
_{acm}=0$, that is, when there is no tendency for the banks’ money creation powers to move in opposite directions. Of course, when both $\mu_{cm}$ and $\mu_{acm}$ are not zero,
the dynamics can be even richer and include (realistically) cases where banks can change behavior over time, from co-moving to moving in different direction, or vice versa.

The Heisenberg equations of motion $\dot{X}(t)=i[H,X(t)]$ can now be used to deduce,
by using the CAR (\ref{21}) and (\ref{23}) and the operator $H$ given in (\ref{24}%
), the following set of differential equations:
\begin{equation}
\left\{
\begin{array}{ll}
\dot{b}_{1}(t)=-i\omega _{1}b_{1}(t)+i\lambda _{1}\int_{\mathbb{R}%
}B_{1}(k,t)\,dk-i\mu_{acm}b_{2}(t)-i\mu_{cm}b_{2}^{\dagger }(t), &  \\
\dot{b}_{2}(t)=-i\omega _{2}b_{2}(t)+i\lambda _{2}\int_{\mathbb{R}%
}B_{2}(k,t)\,dk-i\mu_{acm}b_{1}(t)+i\mu_{cm}b_{1}^{\dagger }(t), &  \\
\dot{B}_{j}(k,t)=-i\Omega _{j}(k)B_{j}(k,t)+i\lambda _{j}b_{j}(t), &
\end{array}%
\right.  \label{25}
\end{equation}%
$j=1,2$. The solution of this system of equations can be found as in
 \cite{Bagarello2015a,Baghavkhr}, and it can be written as follows:
\begin{equation}
b(t)=e^{i\,U\,t}b(0)+i\int_{0}^{t}e^{i\,U\,(t-t_{1})}\,\beta (t_{1})\,dt_{1},
\label{28}
\end{equation}%
where we have introduced the following quantities:
\begin{equation*}
b(t)=\left(
\begin{array}{c}
b_{1}(t) \\
b_{2}(t) \\
b_{1}^{\dagger }(t) \\
b_{2}^{\dagger }(t) \\
\end{array}%
\right) ,\,\beta (t)=\left(
\begin{array}{c}
\lambda _{1}\beta _{1}(t) \\
\lambda _{2}\beta _{2}(t) \\
-\lambda _{1}\beta _{1}^{\dagger }(t) \\
-\lambda _{2}\beta _{2}^{\dagger }(t) \\
\end{array}%
\right) ,\,U=\left(
\begin{array}{cccc}
i\nu _{1} & -\mu_{acm} & 0 & -\mu_{cm} \\
-\mu_{acm} & i\nu _{2} & \mu_{cm} & 0 \\
0 & \mu_{cm} & i\overline{\nu _{1}} & \mu_{acm} \\
-\mu_{cm} & 0 & \mu_{acm} & i\overline{\nu _{2}}%
\end{array}%
\right) ,
\end{equation*}%
and where $\Omega _{j}(k)=\Omega _{j}k$, $\Omega _{j}>0$, $\nu _{j}=i\omega
_{j}+\pi \frac{\lambda _{j}^{2}}{\Omega _{j}}$ and $\beta _{j}(t)=\int_{%
\mathbb{R}}B_{j}(k)e^{-i\Omega _{j}kt}\,dk$, $j=1,2$.

Now, as in \cite{Bagarello2015a,Baghavkhr}, we have to compute the average of the time
evolution of operators $\hat{n}_{j}(t)=b_{j}^{\dagger
}(t)b_{j}(t)$, on a state over the full system $\mathcal{S}$. Again, some technicalities are needed, since the environment lives in an infinite-dimensional Hilbert space. In particular, these states are assumed to be tensor products of vector states for $%
\mathcal{S}_{\mathcal{G}}$ and states on the environment in the following
way: for each operator of the form $X_{\mathcal{S}}\otimes Y_{\mathcal{R}}$,
$X_{\mathcal{S}}$ being an operator of $\mathcal{S}_{\mathcal{G}}$ and $Y_{%
\mathcal{R}}$ an operator of the environment, we have
\begin{equation*}
\left\langle X_{\mathcal{S}}\otimes Y_{\mathcal{R}}\right\rangle
:=\left\langle \Psi _{0},X_{\mathcal{S}}\Psi _{0}\right\rangle \,\omega _{%
\mathcal{R}}(Y_{\mathcal{R}}).
\end{equation*}%
Here $\Psi _{0}$ is the vector introduced in (\ref{20}), while $\omega _{%
\mathcal{R}}(.)$ is a state satisfying the following standard properties,
see \cite{Barnett1997}:
\begin{equation}
\omega _{\mathcal{R}}(1\!\!1_{\mathcal{R}})=1,\quad \omega _{\mathcal{R}%
}(B_{j}(k))=\omega _{\mathcal{R}}(B_{j}^{\dagger }(k))=0,\quad \omega _{%
\mathcal{R}}(B_{j}^{\dagger }(k)B_{l}(q))=N_{j}\,\delta _{j,l}\delta (k-q),
\label{29}
\end{equation}%
for some constant $N_{j}$. 
Also, $\omega _{\mathcal{R}}(B_{j}(k)B_{l}(q))=0$%
, for all $j$ and $l$. We refer to \cite{Bagarello2012,Bagarello2019} for more details on this approach, including states, Hamiltonians and dynamics. { Here ${N}_{j}$ can be considered as the optimal value of $r^D_j$, to which bank j's money creation, $\hat{n}_{j}$, tends to converge dynamically, if the banks do not interact. Note that ${N}_{j}$ is determined by bank j's environment. Obviously, the the actual realizations of  $r^D_j$ over time may differ form the optimal value, ${N}_{j}$, due to the bank's choices (as influenced by its interactions with the environment and the other banks). Such differences expose the bank to (liquidity, credit and settlement) risks. }

After a few computations, calling $V(t)=e^{i\,U\,t}$ and $V_{k,l}(t)$ its $%
(k,l)$-matrix element, we deduce the following general formulas for the
\emph{Quantum functions} (QFs) of $\B_{1}$ and $\B_{2}$,
 \cite{Baghavkhr}:

\begin{equation}
\left\{
\begin{array}{ll}
n_{1}(t)=\left\langle  b_{1}^{\dagger }(t)b_{1}(t) \right\rangle =\mu _{1}^{(\B)}(t)+\delta \mu _{1}^{(\B)}(t)+n_{1}^{(\Bc)}(t), &  \\
n_{2}(t)=\left\langle  b_{2}^{\dagger }(t)b_{2}(t)\right\rangle =\mu _{2}^{(\B)}(t)+\delta \mu _{2}^{(\B)}(t)+n_{2}^{(\Bc)}(t). &
\end{array}%
\right.   \label{230}
\end{equation}%
Here, we have introduced \ \ \
\begin{equation}
\left\{
\begin{array}{ll}
\mu _{1}^{(\B)}(t)=\left\vert V_{1,1}(t)\right\vert ^{2}\left(
\left\vert \alpha _{1,0}\right\vert ^{2}+\left\vert \alpha _{1,1}\right\vert
^{2}\right) +\left\vert V_{1,2}(t)\right\vert ^{2}\left( \left\vert \alpha
_{0,1}\right\vert ^{2}+\left\vert \alpha _{1,1}\right\vert ^{2}\right) + &
\\
\qquad \qquad +\left\vert V_{1,3}(t)\right\vert ^{2}\left( \left\vert \alpha
_{0,0}\right\vert ^{2}+\left\vert \alpha _{0,1}\right\vert ^{2}\right)
+\left\vert V_{1,4}(t)\right\vert ^{2}\left( \left\vert \alpha
_{0,0}\right\vert ^{2}+\left\vert \alpha _{1,0}\right\vert ^{2}\right)  &
\\
\mu _{2}^{(\B)}(t)=\left\vert V_{2,1}(t)\right\vert ^{2}\left(
\left\vert \alpha _{1,0}\right\vert ^{2}+\left\vert \alpha _{1,1}\right\vert
^{2}\right) +\left\vert V_{2,2}(t)\right\vert ^{2}\left( \left\vert \alpha
_{0,1}\right\vert ^{2}+\left\vert \alpha _{1,1}\right\vert ^{2}\right) + &
\\
\qquad \qquad +\left\vert V_{2,3}(t)\right\vert ^{2}\left( \left\vert \alpha
_{0,0}\right\vert ^{2}+\left\vert \alpha _{0,1}\right\vert ^{2}\right)
+\left\vert V_{2,4}(t)\right\vert ^{2}\left( \left\vert \alpha
_{0,0}\right\vert ^{2}+\left\vert \alpha _{1,0}\right\vert ^{2}\right) , &
\end{array}%
\right.   \label{231}
\end{equation}

\begin{equation}
\left\{
\begin{array}{ll}
\delta \mu _{1}^{(\B)}(t)=2\Re \left[ \overline{V_{1,1}(t)}%
\,V_{1,2}(t)\overline{\alpha _{1,0}}\,\alpha _{0,1}+\overline{V_{1,1}(t)}%
\,V_{1,4}(t)\overline{\alpha _{1,1}}\,\alpha _{0,0}\right] + &  \\
\qquad \qquad -2\Re \left[ \overline{V_{1,2}(t)}\,V_{1,3}(t)\overline{\alpha
_{1,1}}\,\alpha _{0,0}+\overline{V_{1,3}(t)}\,V_{1,4}(t)\overline{\alpha
_{0,1}}\,\alpha _{1,0}\right] , &  \\
\delta \mu _{2}^{(\B)}(t)=2\Re \left[ \overline{V_{2,1}(t)}%
\,V_{2,2}(t)\overline{\alpha _{1,0}}\,\alpha _{0,1}+\overline{V_{2,1}(t)}%
\,V_{2,4}(t)\overline{\alpha _{1,1}}\,\alpha _{0,0}\right] + &  \\
\qquad \qquad -2\Re \left[ \overline{V_{2,2}(t)}\,V_{2,3}(t)\overline{\alpha
_{1,1}}\,\alpha _{0,0}+\overline{V_{2,3}(t)}\,V_{2,4}(t)\overline{\alpha
_{0,1}}\,\alpha _{1,0}\right] , &
\end{array}%
\right.   \label{232}
\end{equation}%
and
\begin{equation}
\left\{
\begin{array}{ll}
n_{1}^{(\Bc)}(t)=2\pi \int_{0}^{t}dt_{1}\left[ \frac{\lambda _{1}^{2}}{\Omega
_{1}}\left(
|V_{1,1}(t-t_{1})|^{2}N_{1}+|V_{1,3}(t-t_{1})|^{2}(1-N_{1})\right) \right] +
&  \\
\qquad \qquad +2\pi \int_{0}^{t}dt_{1}\left[ \frac{\lambda _{2}^{2}}{\Omega
_{2}}\left(
|V_{1,2}(t-t_{1})|^{2}N_{2}+|V_{1,4}(t-t_{1})|^{2}(1-N_{2})\right) \right] ,
&  \\
n_{2}^{(\Bc)}(t)=2\pi \int_{0}^{t}dt_{1}\left[ \frac{\lambda _{1}^{2}}{\Omega
_{1}}\left(
|V_{2,1}(t-t_{1})|^{2}N_{1}+|V_{2,3}(t-t_{1})|^{2}(1-N_{1})\right) \right] +
&  \\
\qquad \qquad +2\pi \int_{0}^{t}dt_{1}\left[ \frac{\lambda _{2}^{2}}{\Omega
_{2}}\left(
|V_{2,2}(t-t_{1})|^{2}N_{2}+|V_{2,4}(t-t_{1})|^{2}(1-N_{2})\right) \right] .
&
\end{array}%
\right.   \label{233}
\end{equation}%
In formula (\ref{230}) we have clearly divided contributions of three
different natures: $\mu _{j}^{(\B)}(t)$ contains contributions only
due to the banks. 
$\delta \mu _{1}^{(\B)}(t)$ and $%
\delta \mu _{2}^{(\B)}(t)$ also depend only on the banks. They are \emph{interference terms}, since they can be different from zero only if the initial state of the bank, i.e. the vector $\Psi_0$ in (\ref{20}), is some non trivial linear combination. If, on the other hand, $\Psi_0$ coincides with one of the $\varphi_{k,l}$, this implies that all coefficients $\alpha_{k,l}$ in (\ref{20}) are zero, except one. Then, $\delta \mu _{1}^{(\B)}(t) = \delta \mu _{2}^{(\B)}(t)=0$. Hence, these terms contribute to $n_j(t)$ only if the initial state of the bank is not {\em sharp}, i.e., if there is uncertainty on their state at $t=0$. This uncertainty creates the interference. Finally, $%
n_{1}^{(\Bc)}(t)$ and $n_{2}^{(\Bc)}(t)$ arise because of the interaction of the
banks with their environments: as we see from (\ref{233}), they are both
zero if $\lambda _{1}$ and $\lambda _{2}$ in the Hamiltonian are equal
to zero, and they do not depend on the explicit form of $\Psi _{0}$.

Of course, the first two terms simplify significantly if the initial state $\Psi_0$ is not a linear combination of $\varphi_{k,l}$, but it is just one of these vectors; that is, if the initial state of the banks is unambiguosly defined and there is no uncertainty. This is because all coefficients $%
\alpha _{k,l}$ in (\ref{20}) are zero, except one.  In this particular
situation, we observe that  $\delta \mu _{j}^{(\B)}(t)=0$, $j=1,2$.

\section{Some numerical simulations}\label{sectSNS}

In this section, we focus on the role of the various parameters in the analysis of our financial system. For that, we consider three different scenarios. In the first one, the two banks interact with their environments, but not among themselves. This corresponds to taking $H_{int}=0$ in (\ref{24}). In the second scenario $H_I=0$; here, the two banks interact with one another but not with their environments. 
The third (and last) scenario is the one in which all the interactions are active: both $H_{int}$ and $H_I$ are non zero. Various other simulations are reported in Appendix B.

Before starting our analysis, it is useful to spend some words on our specific choice of the model's parameters. Indeed, one may imagine that their values have been chosen looking for the best fit of some experimental data. This is the usual approach in numerics. However, we have no experimental data to fit, at least in this article, but only qualitative behaviors we want to describe using our model. At this stage, we are much more interested in seeing the analytical gains that can be obtained by applying quantum mechanics to banks' deposit liabilities as hybrid instruments (part debt, and part equity), in line with the Accounting View of Money. Fitting numbers would be an undertaking for a later stage, and the complexity of the exercise would require, we believe, a refinement of the model proposed here, using, for instance, $(H,\rho)$-induced dynamics, \cite{bagrules}. For this reason, it should be clear that the explicit numerical value of the parameters used in the simulations below is not so relevant for the purpose of this article. They have been chosen to adapt the quantum model to the problem at hand and to make it tractable in an economic context. It is rather more relevant for our purposes to consider the relation between different parameters. This is particularly the case in the simple system described in Section \ref{sectpredprey}. For instance, if we take $\omega_1$ and $\omega_2$ to differ significantly from one another, formulas in (\ref{MM26-27}) show that $n_1(t)$ and $n_2(t)$ have small oscillations around their mean values. The amplitude of these oscillations increases when $\omega_1$ approaches $\omega_2$. As shown below, this is a central result of the quantum model as applied to bank money: the more divergent the money supply behavior of interacting banks, the more stable their debt/equity deposit mix over time and, thus, the lower their exposure to (liquidity, credit and settlement) risk. Going back to our model, it is maybe useful to observe that, because of the fermionic nature of our operators, there are natural constraints which must be satisfied. For instance, in what follows, $N_1$ and $N_2$ take values in the interval $[0,1]$, being connected to reservoirs of fermionic nature - or the bank's environments, in economic terms, which, as noted earlier determine $N_1$ and $N_2$ as optimal values of each bank's debt/equity deposit mix (i.e., $r^D_j$).

\subsection{Scenario 1: No interactions between $\B_{1}$ and $\B_{2}$}

\begin{figure}[th]
	\begin{center}
		\includegraphics[width=0.4\textwidth]{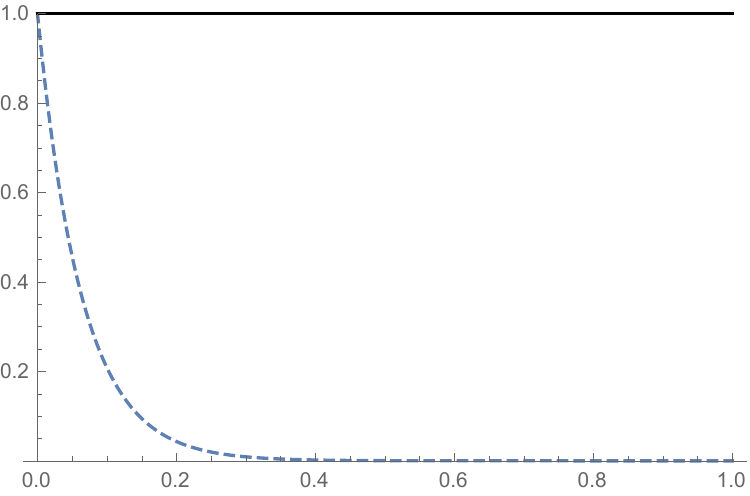}\hspace{%
			8mm} \includegraphics[width=0.4\textwidth]{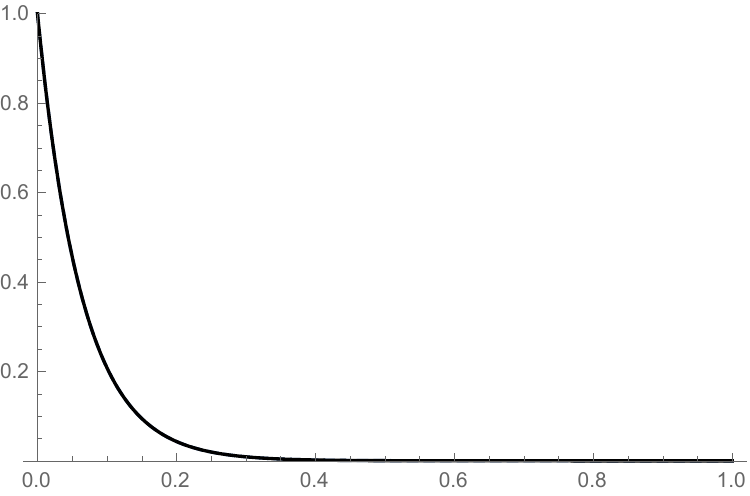}%
		\hfill\\[8pt]
		\includegraphics[width=0.4\textwidth]{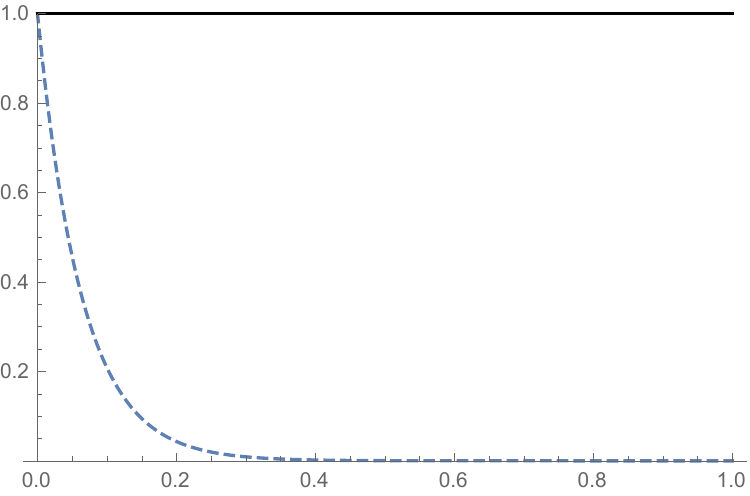}\hspace{%
			8mm} \includegraphics[width=0.4\textwidth]{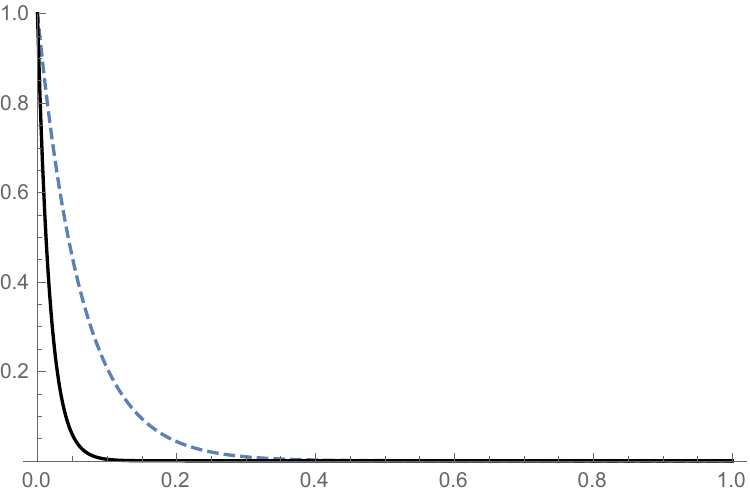}%
		\hfill\\[8pt]
				\includegraphics[width=0.4\textwidth]{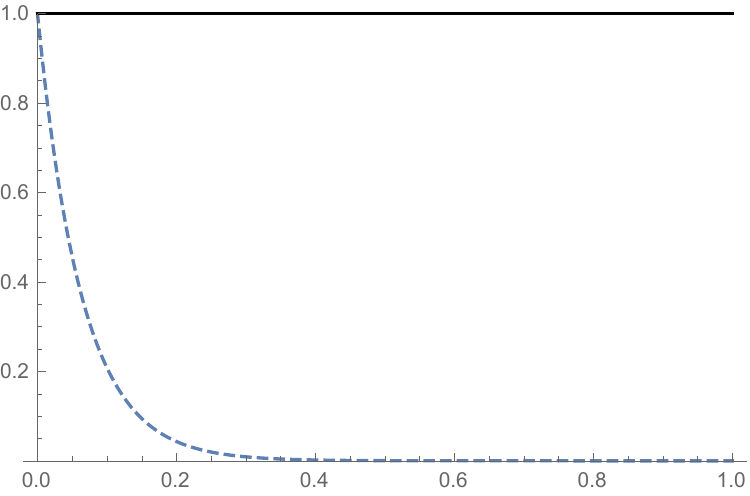}\hspace{%
			8mm} \includegraphics[width=0.4\textwidth]{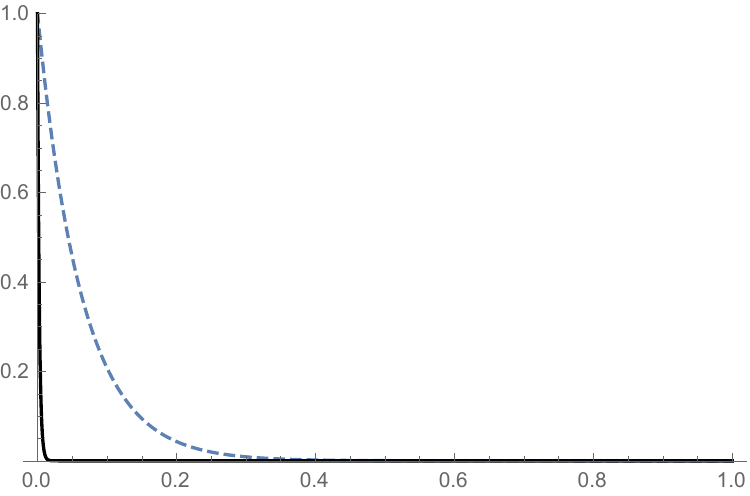}%
		\hfill\\[8pt]
		\includegraphics[width=0.4\textwidth]{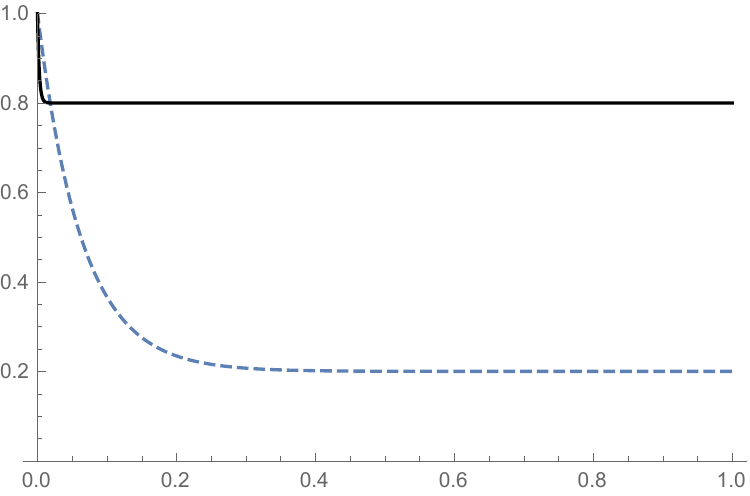}\hspace{%
			8mm} \includegraphics[width=0.4\textwidth]{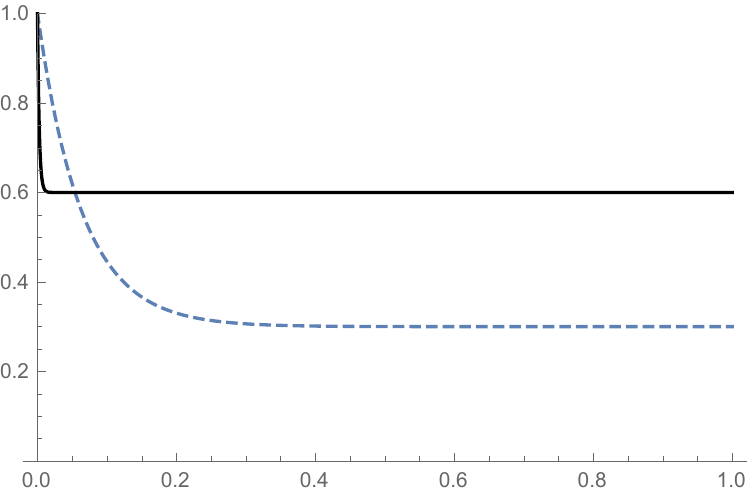}%
	\end{center}
	\caption{{\protect\footnotesize The QFs $n_{1}(t)$ (dotted line) and $n_{2}(t)$
			(continuous line) for $\mu_{acm}=\mu_{cm}=0$. The other parameters are given in the text. }}
	\label{fig7}
\end{figure}

We start our analysis by putting $\mu_{cm}=\mu_{acm}=0$, that there is no systematic patterns of the way in which the money creation (deposit issuance) of the two banks moves. Hence $H_{int}=0$, and the whole system is decoupled into two subsystems, one, $\F_1$, including $\B_1$ and $B_1$, and the other, $\F_2$, $\B_2$ and $B_2$. For this reason, it is sufficient to change the parameters of, say $\F_2$, while keeping fixed those of $\F_1$, to analyze their role and meaning for our system. In Figure \ref{fig7} we show $n_1(t)$ and $n_2(t)$ for the following set of parameters and initial conditions: we assume that $\Psi_0=\varphi_{1,1}$, so that the only non zero $\alpha_{k,l}$ in (\ref{20}) is $\alpha_{1,1}=1$. We also take $\omega_1=1$, $\Omega_1=.1$ and $\lambda_1=.5$ for $\F_1$ (dotted lines), always fixed. The analogous parameters for $\F_2$ (continuous lines) are different in each row: $\omega_2=2$, $\Omega_2=.1$ and $\lambda_2=.5$ in the first row, $\omega_2=2$, $\Omega_2=1$ and $\lambda_2=3$ in the second row, $\omega_2=2$, $\Omega_2=1$ and $\lambda_2=8$ in the third and in the fourth rows. In the left column, row 1, 2 and 3, we have $N_1=0$ and $N_2=1$, while, in the fourth row, $N_1=0.2$ and $N_2=0.8$. In the right column the first three rows have $N_1=N_2=0$, while, in the fourth row, $N_1=0.3$ and $N_2=0.6$.

These plots show that, in absence of interactions between $\B_1$ and $\B_2$, the model produces the following outputs, which could also be deduced with different choices of parameters and initial conditions:

\begin{enumerate}
	
	\item the QFs behave monotonically. They always decrease, or stay constant (however, changing $\Psi_0$ we could also get increasing QFs), and the limit is always the value of $N_1$ and $N_2$. The final values of the QFs is only fixed by their environments, that is the behavior of the two banks is determined only by their local conditions.
	
	\item The role of the other parameters in $H$ is relevant to determine the {\em speed of reaction} of the banks. In particular, while the dotted lines $n_1(t)$ in Figure \ref{fig7} decay (to $N_1=0$) always with the same speed, the function $n_2(t)$ decays to its limiting value ($N_2=1$) faster and faster when $\Omega_2$ and/or $\lambda_2$ increase; that is, when the contribution of each bank's environment to its money creation power and/or each bank's interactions with its environments increases, the function's convergence to environment-determined $N_2=1$ is more rapid. On the other hand, increasing $\omega_2$ would, in general, slower this procedure, as discussed in several other systems, \cite{Bagarello2012,Bagarello2019}; this means that, as the marginal cost of the bank to change its money creation rises, the function's dynamics would slow down. 
	
\end{enumerate}

Summarizing: each bank behaves in a way which is suggested by its own environment, and its adjustment is monotonous. However, the {\em efficiency} of the change depends on the other parameters of $H$.

\subsection{Scenario 2: No interactions between $\B_{j}$ and $B_{j}$}

\begin{figure}[th]
	\begin{center}
		\includegraphics[width=0.4\textwidth]{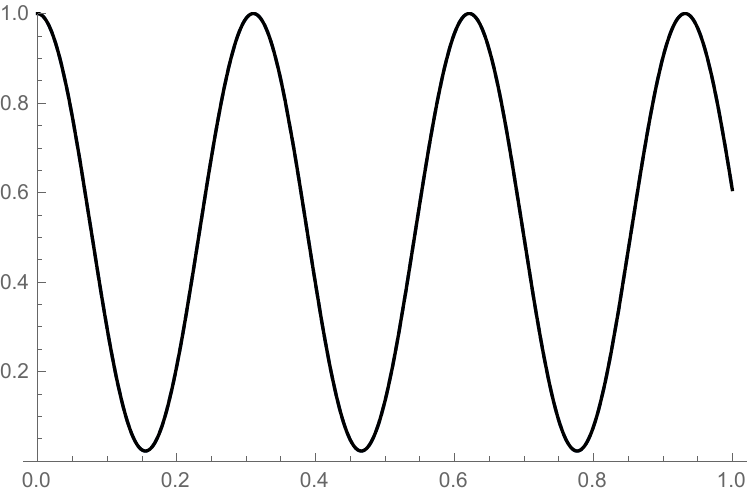}\hspace{%
			8mm} \includegraphics[width=0.4\textwidth]{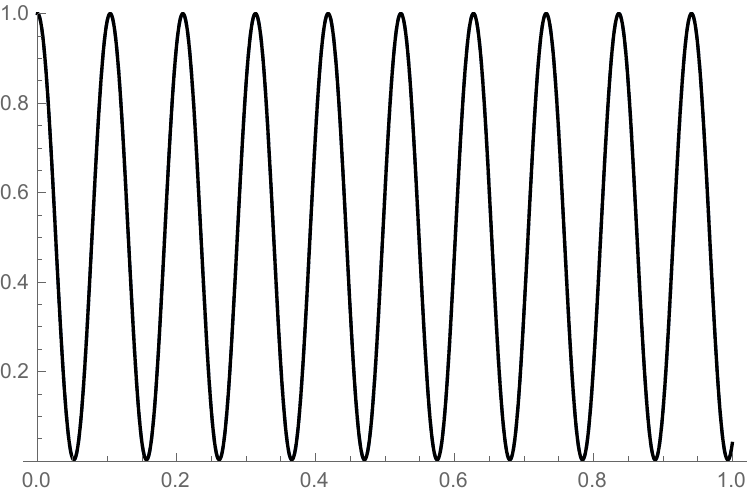}%
		\hfill\\[10pt]
		\includegraphics[width=0.4\textwidth]{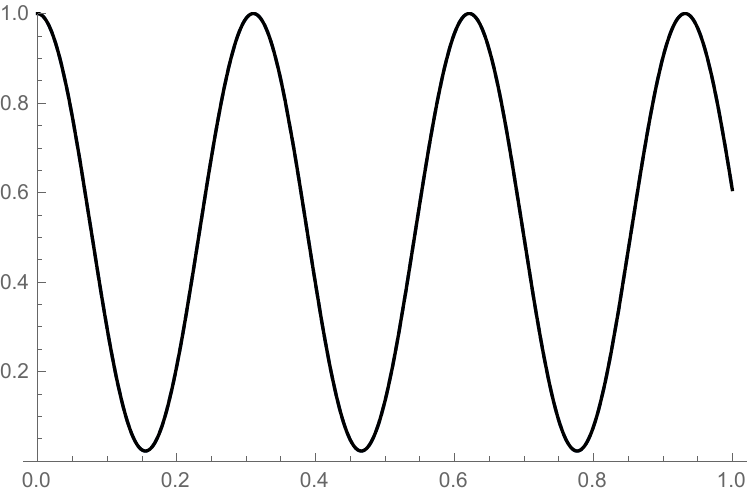}\hspace{%
			8mm} \includegraphics[width=0.4\textwidth]{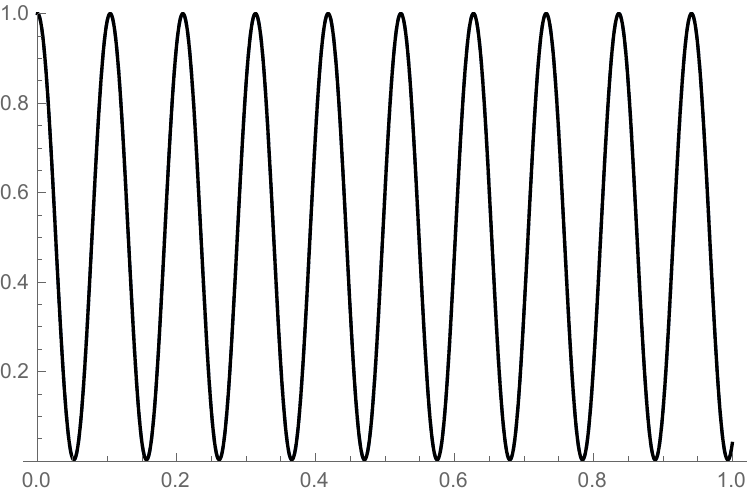}%
	\end{center}
	\caption{{\protect\footnotesize The QFs $n_{1}(t)$ (dotted line) and $n_{2}(t)$
			(continuous line) for $\lambda_1=\lambda_2=0$, $\omega_1=1$, $\omega_2$ and $\Psi_0=\varphi_{1,1}$. Moreover we have: top left $\mu_{acm}=100$, $\mu_{cm}=10$; top right $\mu_{acm}=100$, $\mu_{cm}=30$; down left $\mu_{acm}=0$, $\mu_{cm}=10$; down right $\mu_{acm}=0$, $\mu_{cm}=30$.}}
	\label{fig8}
\end{figure}

\begin{figure}[th]
	\begin{center}
		\includegraphics[width=0.4\textwidth]{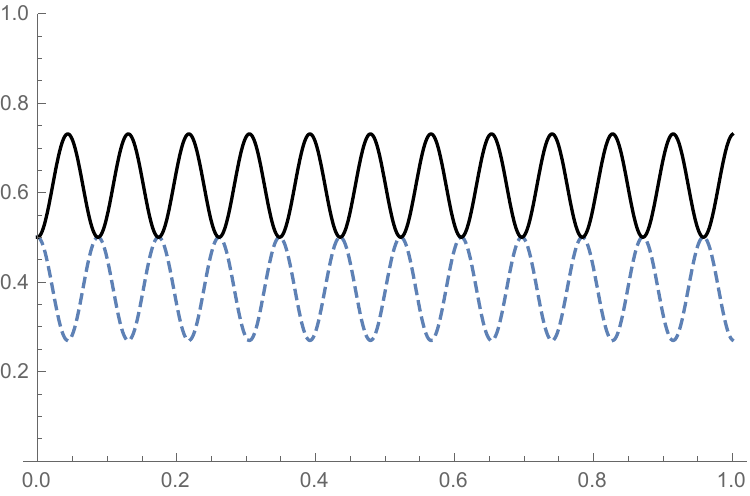}\hspace{%
			8mm} \includegraphics[width=0.4\textwidth]{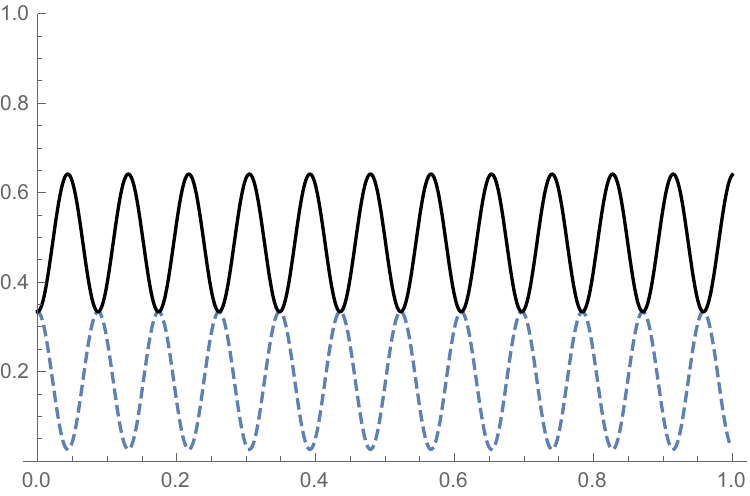}%
	\end{center}
	\caption{{\protect\footnotesize The QFs $n_{1}(t)$ (dotted line) and $n_{2}(t)$
			(continuous line) for $\lambda_1=\lambda_2=0$, $\omega_1=20$, $\omega_2=60$, $\mu_{acm}=30$ and $\mu_{cm}=0$. $\Psi_0$ is given in the text.}}
	\label{fig9}
\end{figure}

In this case, we put $\lambda_1=\lambda_2=0$ in (\ref{24}). This means that $H_I=0$. In this case, the two banks do not entertain any interaction with their environments. From a mathematical point of view, the system lives in a Hilbert space which, now, corresponds to $\Hil_{\B}$, which is four-dimensional. It is well known, see \cite{Bagarello2019} for instance, that in this case the motion of the system driven by an Hermitian Hamiltonian (like ours, independently of any detail!), is necessarily periodic or quasi-periodic. This is exactly what we observe in Figure \ref{fig8}: the QFs keep on oscillating (or stay constant), which means that $r^D_j$ is varying all the time (or exhibits trivial dynamics). This is because the environment-determined $N_j's$ do not exert any attractive force on the QFs. Of course, if other than having $\lambda_1=\lambda_2=0$ we also consider $\mu_{cm}=\mu_{acm}=0$, the situation trivializes since $H$ consists only of the free part: no interaction and, therefore, no interesting dynamics. For this reason, we are not interested in this simple case and only consider more realistic cases where $\mu_{cm}$ or $\mu_{acm}$, or both, are different from zero. In particular, the plots in Figure \ref{fig8} and \ref{fig9} suggest that the role of $\mu_{cm}$ is, in a sense, more relevant than that of $\mu_{acm}$: it is sufficient to have $\mu_{cm}\neq0$ for the system to show oscillations independently of the choice of $\Psi_0$. This suggests that, all else being equal, banks increasing or decreasing their money creation together cause the debt/equity deposit mix to oscillate over time, exposing the banks to higher (liquidity, credit and settlement) risks. On the other hand, the only way to get a non trivial oscillating behavior of the QFs if $\mu_{cm}=0$ and $\mu_{acm}\neq0$, is that $\Psi_0$ is a combination of the different $\varphi_{j,l}$. In fact, what we observe in our simulations, is that if $\Psi_0=\varphi_{j,l}$ for some $j$ and $l$, then having $\mu_{cm}=0$ and $\mu_{acm}\neq0$ leaves the system in a stationary state. This can be understood because the co-movement term in $H_{int}$ {\em amplifies} the effect of the interactions between $\B_1$ and $\B_2$, while the amti-comovement term tends to cancel this effect out, due to the joint presence of a raising and a lowering operator.

It should be clarified that in these plots the functions $n_1(t)$ and $n_2(t)$ simply overlap. We should also stress that the choice of $\Omega_1$, $\Omega_2$, $N_1$ and $N_2$ is unessential here, since the fact that  $\lambda_1=\lambda_2=0$ implies that the environments simply don't contribute to the dynamics. Comparing the first and the second rows in Figure \ref{fig8} it is evident that the oscillations are only connected to the presence of a non zero $\mu_{cm}$, and their frequencies increase with $\mu_{cm}$, but not with $\mu_{acm}$, as noted above.

In Figure \ref{fig9} we consider two different choices of the initial state. In the left plot we have $\alpha_{k,l}=\frac{1}{2}$, $k,l=0,1$, in (\ref{20}). In the right plot we have $\alpha_{0,0}=\alpha_{1,0}=\alpha_{0,1}=\frac{1}{\sqrt{3}}$, while $\alpha_{1,1}=0$. These plots are interesting since they show that, if $\Psi_0$ is some suitable mixture of the $\varphi_{k,l}$, then oscillations of the QFs can occur also when $\mu_{cm}=0$. This is not what we observe in our numerical situations when $\Psi_0$ is described by a single $\varphi_{k,l}$, for some fixed $k$ and $l$, under the same condition on $\mu_{cm}=0$. In this case, we always find $n_1(t)=k$ and $n_2(t)=l$, independently of $t$: we need non trivial linear combinations for $\Psi_0$ if we want to have some oscillations when $\mu_{cm}=0$.

\subsection{Scenario 3: The effect of the whole Hamiltonian}

In this section, we consider what happens when all the terms of $H$ in (\ref{24}) are non zero. We have already considered in Section IV what happens when one of $\mu_{acm}$ and $\mu_{cm}$ is zero. Here, we focus on the more general case where both these parameters are non zero, which is what typically happens in the real world as banks may change behavior over time, i.e., their money supply dynamics may co-move at times (and given certain conditions) or may move in opposite directions or independently of one another. It is useless to notice that there very many different choices are possible of the numerical values of the parameters of the model. Here we concentrate on some of them, which we think are particularly interesting, and we draw our conclusions based on them. Other choices have also been considered, but they do not affect our conclusions. Also, it is not hard to see that the results are stable under small perturbations: changing the values of the parameters by some small amount produces small changes in the plots of the QFs. This is not surprising, since our equations in (\ref{25}) are linear.

We have considered here four different choices of $\Psi_0$. The first choice, $\Psi_0^{(1)}$, has $\alpha_{k,l}=1$ if $k=l=1$, and $\alpha_{k,l}=0$ otherwise: $\alpha_{k,l}=\delta_{k,1}\delta_{l,1}$. The second choice, $\Psi_0^{(2)}$, has $\alpha_{k,l}=\frac{1}{2}$ for $k,l=0,1$. The third choice, $\Psi_0^{(3)}$, has $\alpha_{0,0}=-\alpha_{0,1}=\frac{1}{2}$ and $\alpha_{1,0}=-\alpha_{1,1}=\frac{i}{2}$.
 The fourth choice, $\Psi_0^{(4)}$, has $\alpha_{0,0}=\alpha_{1,0}=\alpha_{0,1}=\frac{1}{\sqrt{3}}$ and $\alpha_{1,1}=0$. With these choices we cover several different situations. Notice that  these vectors are all normalized: $\|\Psi_0^{(j)}\|=1$, $j=1,2,3,4$.
  
All the plots in this section have been obtained with the following set of parameters: $\omega_1=1$, $\omega_2=2$, $\Omega_1=\Omega_2=0.1$, $\lambda_1=0.2$ and $\lambda_2=0.3$. We see that $\B_2$ has all values larger than those of $\B_1$. Thus, we can say that the second bank enjoys greater EoS, and thus incurs lower cost of changing money creation, than the first one. In the various figures, we plot the QFs for different values of $(\mu_{acm},\mu_{cm})$, and different values of $(N_1,N_2)$. In this way, we can analyse the effect of the interactions between the banks, and some of the effects of the interactions between each bank and its own environment.

\begin{figure}[th]
	\begin{center}
		\includegraphics[width=0.4\textwidth]{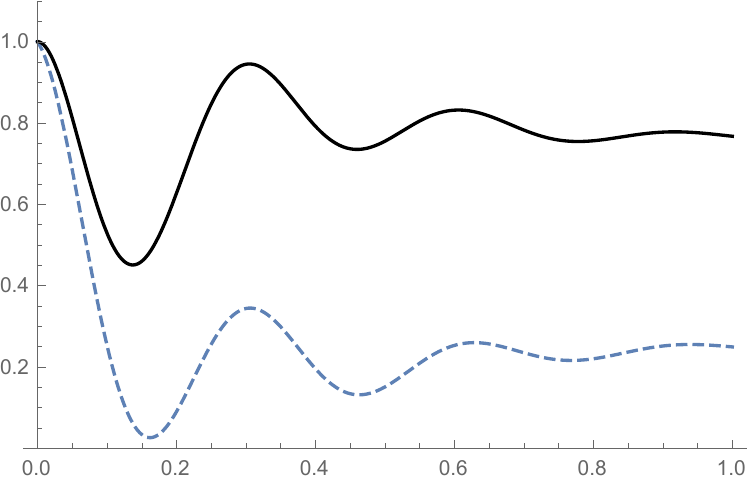}\hspace{%
			8mm} \includegraphics[width=0.4\textwidth]{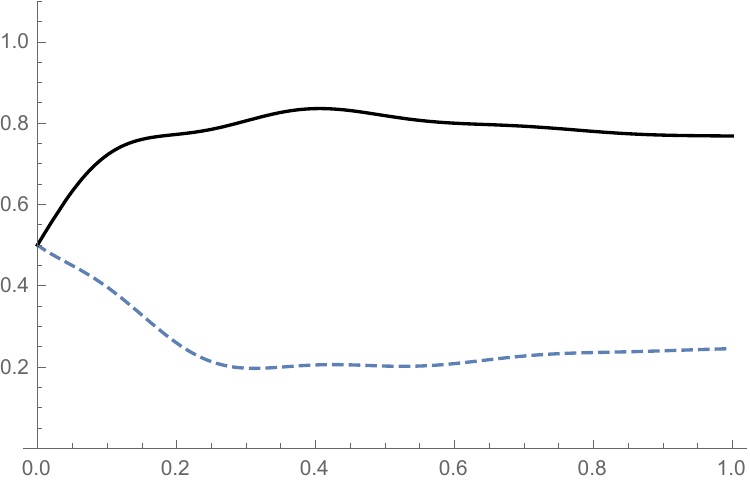}%
		\hfill\\[2pt]
		\includegraphics[width=0.4\textwidth]{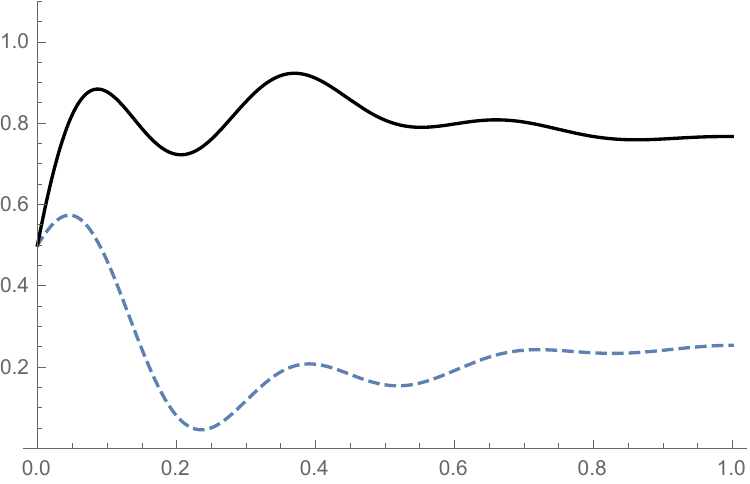}\hspace{%
			8mm} \includegraphics[width=0.4\textwidth]{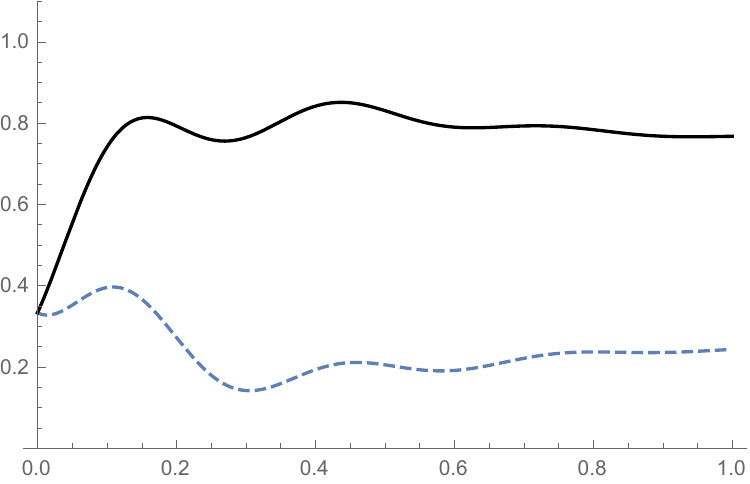}%
	\end{center}
	\caption{{\protect\footnotesize The QFs $n_{1}(t)$ (dotted line) and $n_{2}(t)$
			(continuous line) for $\mu_{acm}=2$, $\mu_{cm}=10$, $N_1=0$, $N_2=1$ and the other parameters given in the text. Initial conditions: top left $\Psi_0^{(1)}$; top right $\Psi_0^{(2)}$; down left $\Psi_0^{(3)}$; down right $\Psi_0^{(4)}$.}}
	\label{fig10}
\end{figure}

\begin{figure}[th]
	\begin{center}
		\includegraphics[width=0.4\textwidth]{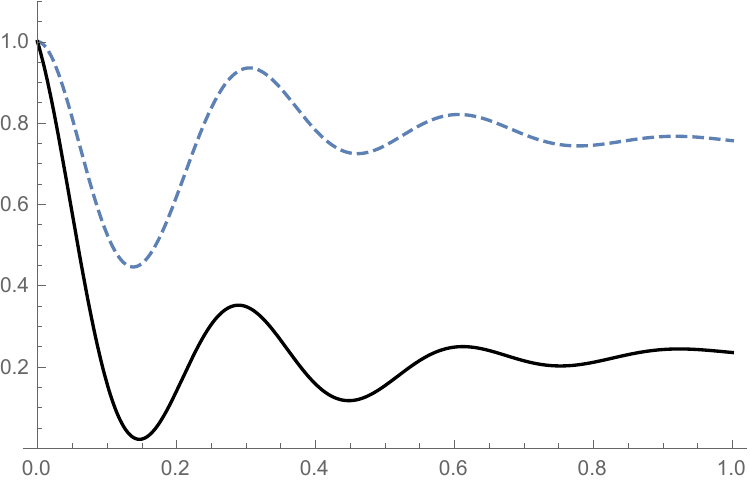}\hspace{%
			8mm} \includegraphics[width=0.4\textwidth]{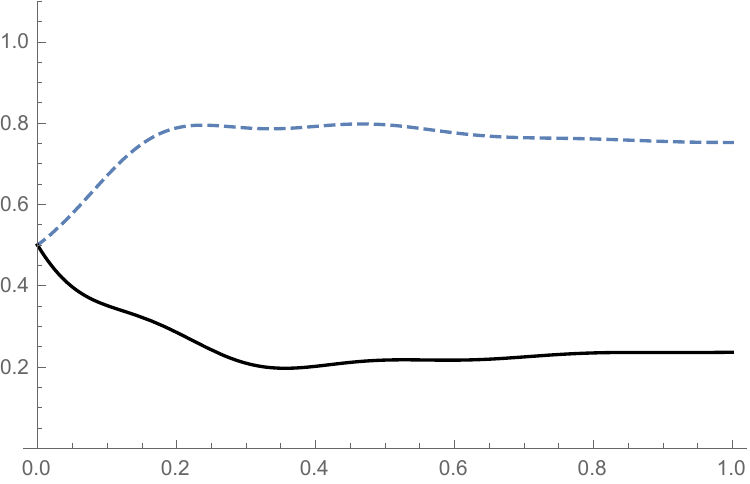}%
		\hfill\\[2pt]
		\includegraphics[width=0.4\textwidth]{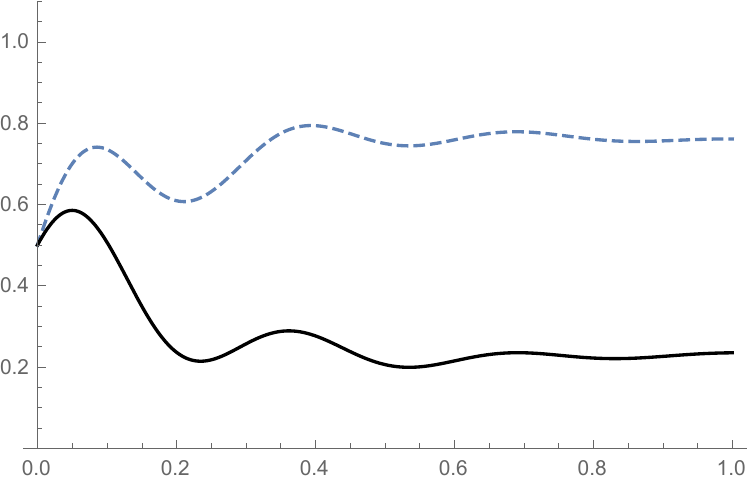}\hspace{%
			8mm} \includegraphics[width=0.4\textwidth]{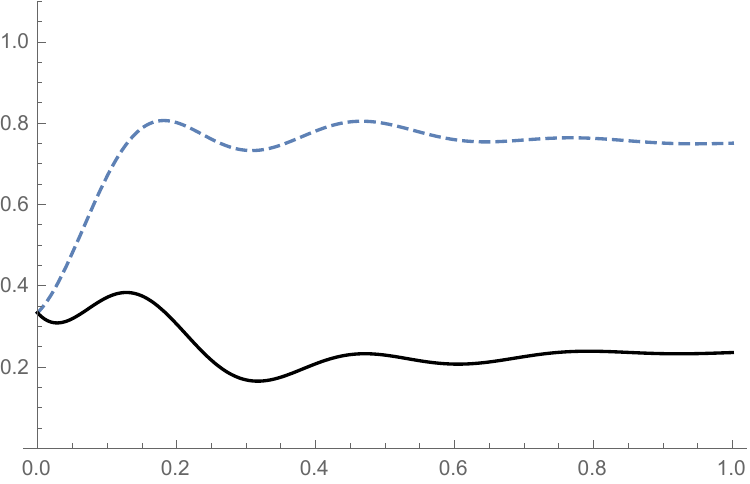}%
	\end{center}
	\caption{{\protect\footnotesize  The QFs $n_{1}(t)$ (dotted line) and $n_{2}(t)$
			(continuous line) for $\mu_{acm}=2$, $\mu_{cm}=10$, $N_1=1$, $N_2=0$ and the other parameters given in the text. Initial conditions: top left $\Psi_0^{(1)}$; top right $\Psi_0^{(2)}$; down left $\Psi_0^{(3)}$; down right $\Psi_0^{(4)}$.}}
	\label{fig11}
\end{figure}

\begin{figure}[th]
	\begin{center}
		\includegraphics[width=0.4\textwidth]{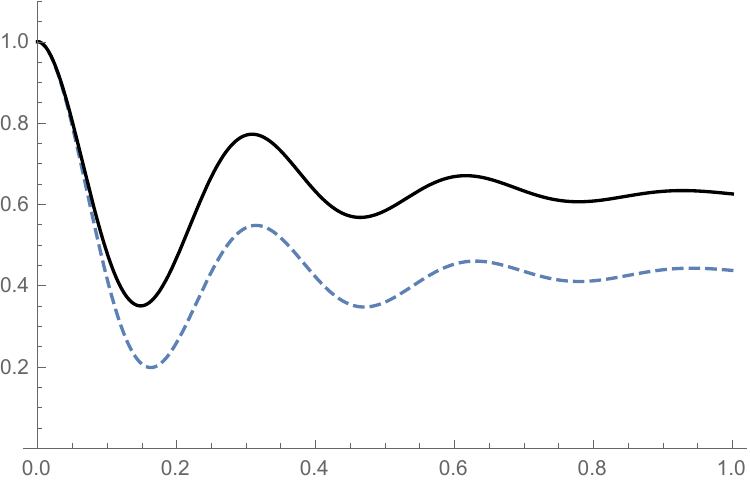}\hspace{%
			8mm} \includegraphics[width=0.4\textwidth]{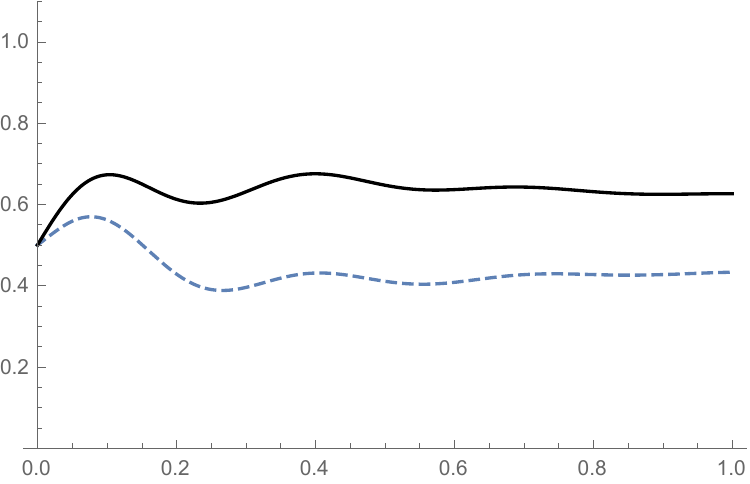}%
		\hfill\\[2pt]
		\includegraphics[width=0.4\textwidth]{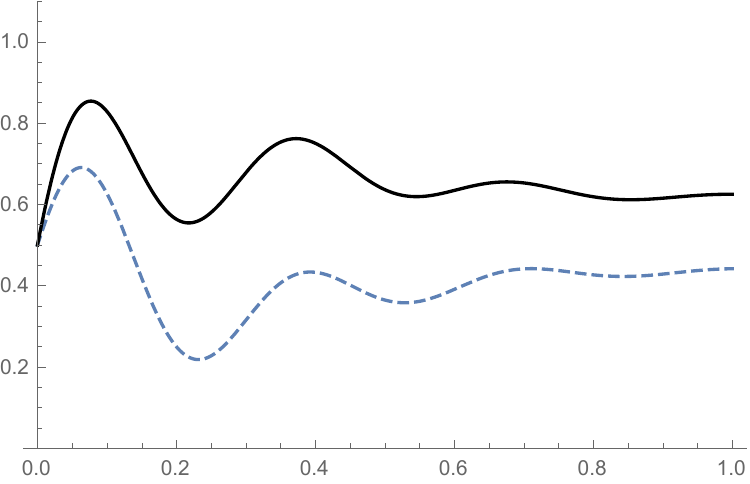}\hspace{%
			8mm} \includegraphics[width=0.4\textwidth]{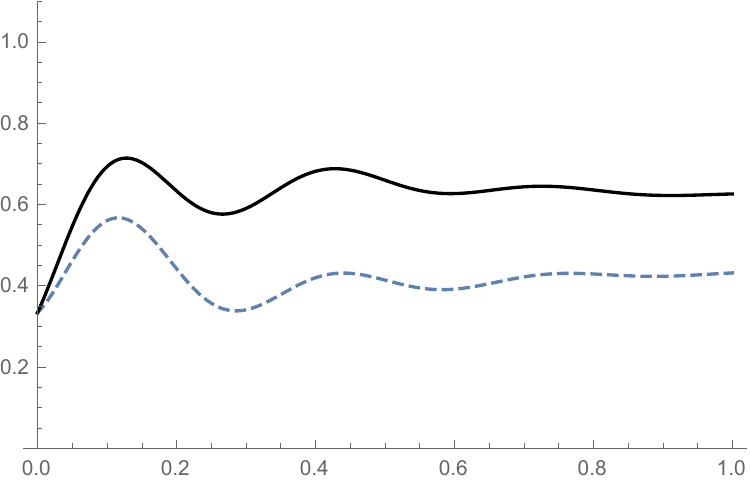}%
	\end{center}
	\caption{{\protect\footnotesize  The QFs $n_{1}(t)$ (dotted line) and $n_{2}(t)$
			(continuous line) for $\mu_{acm}=2$, $\mu_{cm}=10$, $N_1=1$, $N_2=1$ and the other parameters given in the text. Initial conditions: top left $\Psi_0^{(1)}$; top right $\Psi_0^{(2)}$; down left $\Psi_0^{(3)}$; down right $\Psi_0^{(4)}$.}}
	\label{fig12}
\end{figure}

\begin{figure}[th]
	\begin{center}
		\includegraphics[width=0.4\textwidth]{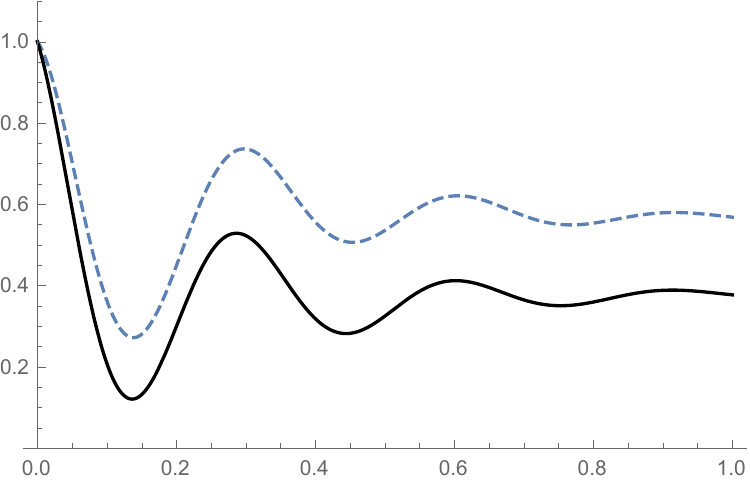}\hspace{%
			8mm} \includegraphics[width=0.4\textwidth]{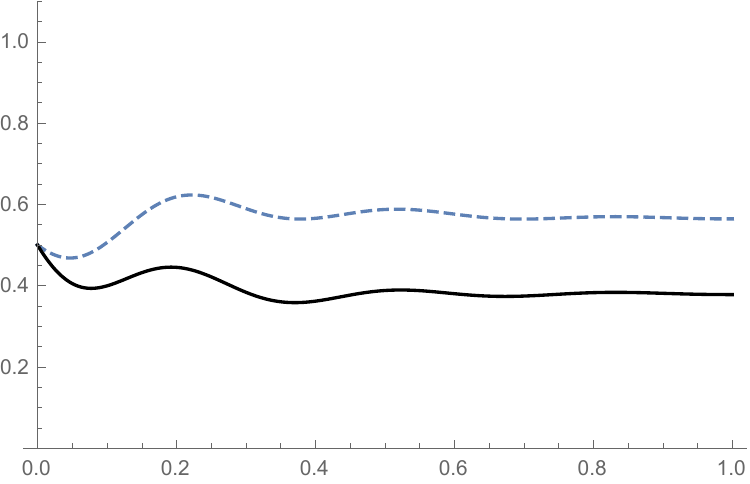}%
		\hfill\\[2pt]
		\includegraphics[width=0.4\textwidth]{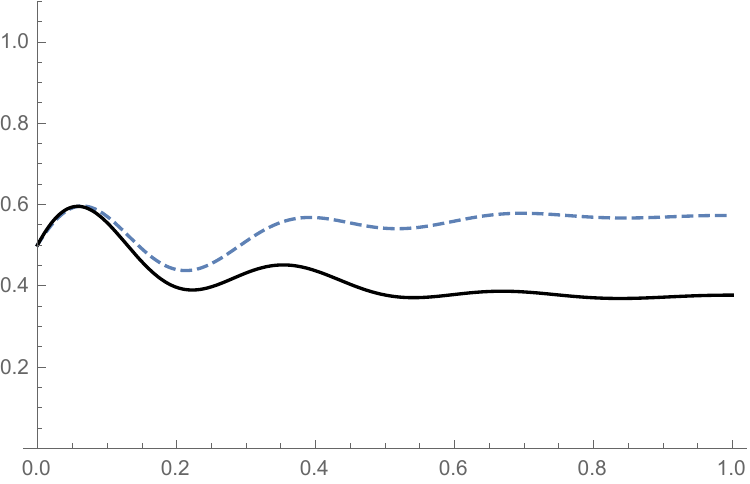}\hspace{%
			8mm} \includegraphics[width=0.4\textwidth]{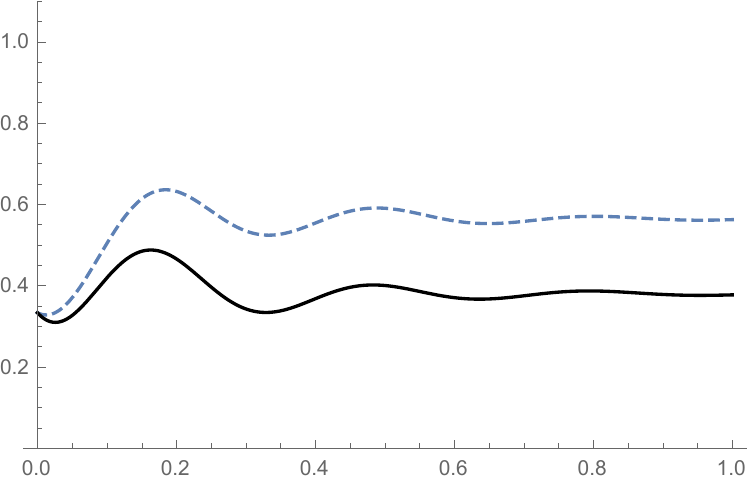}%
	\end{center}
	\caption{{\protect\footnotesize  The QFs $n_{1}(t)$ (dotted line) and $n_{2}(t)$
			(continuous line) for $\mu_{acm}=2$, $\mu_{cm}=10$, $N_1=0$, $N_2=0$ and the other parameters given in the text. Initial conditions: top left $\Psi_0^{(1)}$; top right $\Psi_0^{(2)}$; down left $\Psi_0^{(3)}$; down right $\Psi_0^{(4)}$.}}
	\label{fig13}
\end{figure}

\begin{figure}[th]
	\begin{center}
		\includegraphics[width=0.4\textwidth]{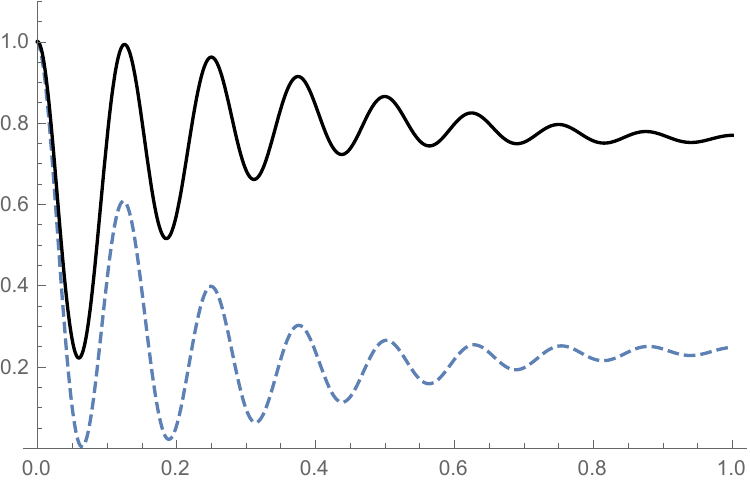}\hspace{%
			8mm} \includegraphics[width=0.4\textwidth]{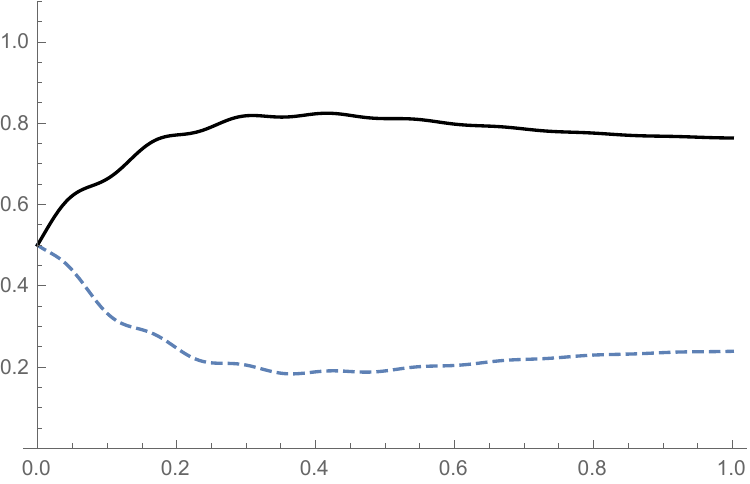}%
		\hfill\\[2pt]
		\includegraphics[width=0.4\textwidth]{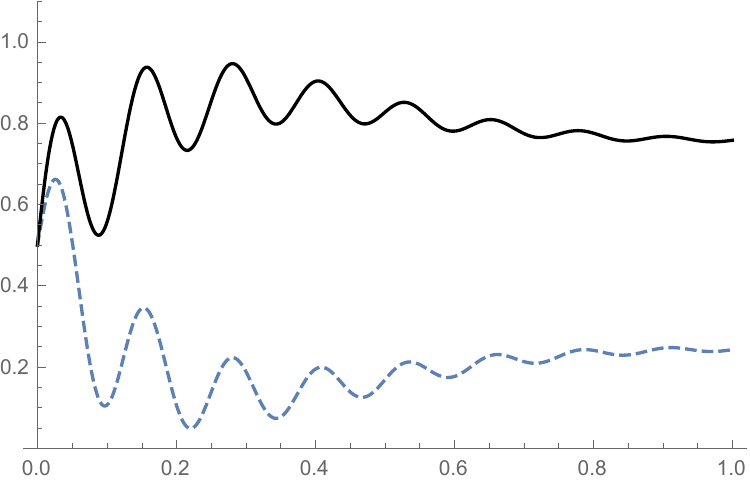}\hspace{%
			8mm} \includegraphics[width=0.4\textwidth]{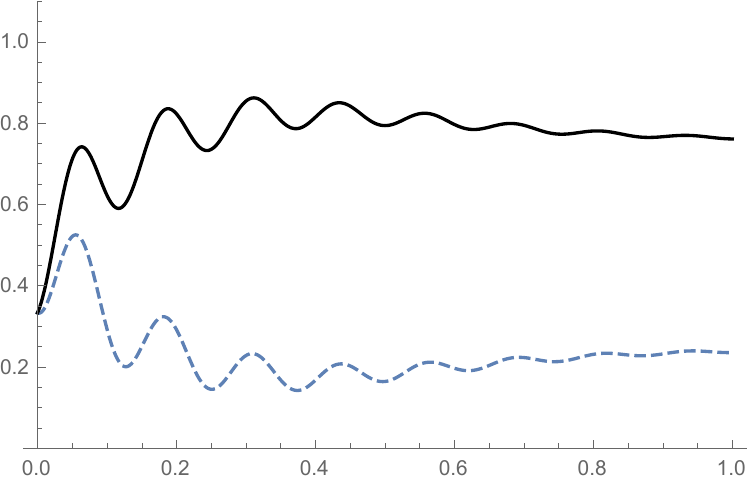}%
	\end{center}
	\caption{{\protect\footnotesize The QFs $n_{1}(t)$ (dotted line) and $n_{2}(t)$
			(continuous line) for $\mu_{acm}=2$, $\mu_{cm}=25$, $N_1=0$, $N_2=1$ and the other parameters given in the text. Initial conditions: top left $\Psi_0^{(1)}$; top right $\Psi_0^{(2)}$; down left $\Psi_0^{(3)}$; down right $\Psi_0^{(4)}$.}}
	\label{fig14}
\end{figure}

\begin{figure}[th]
	\begin{center}
		\includegraphics[width=0.4\textwidth]{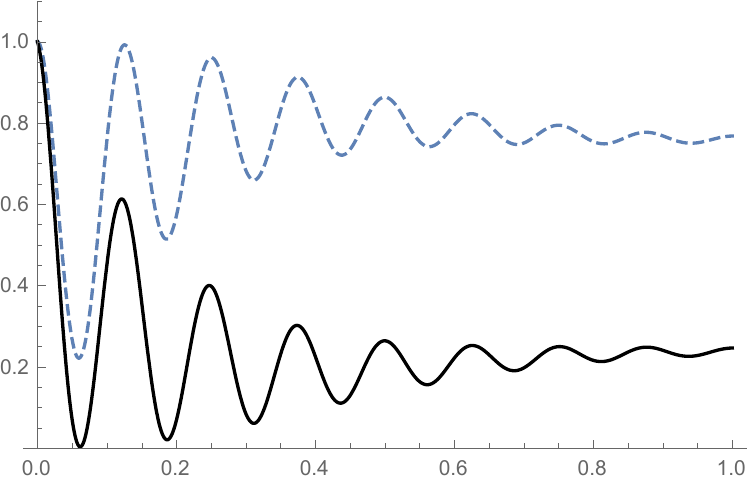}\hspace{%
			8mm} \includegraphics[width=0.4\textwidth]{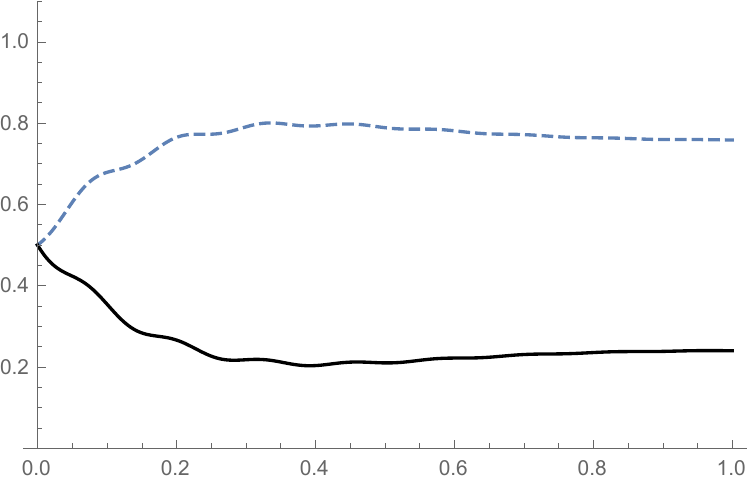}%
		\hfill\\[2pt]
		\includegraphics[width=0.4\textwidth]{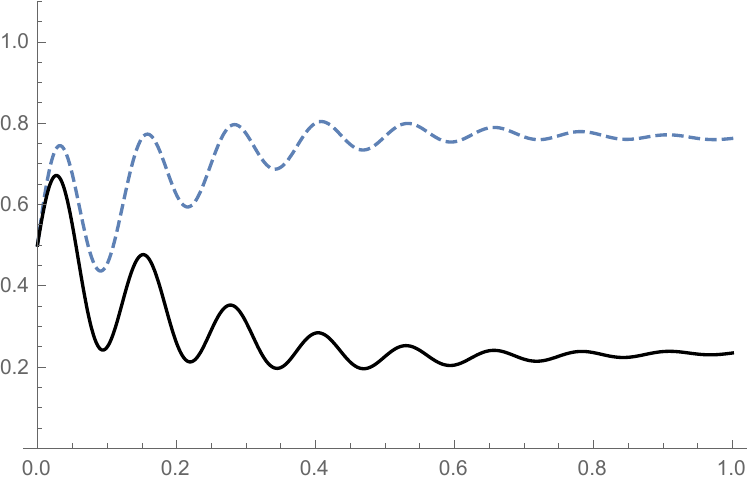}\hspace{%
			8mm} \includegraphics[width=0.4\textwidth]{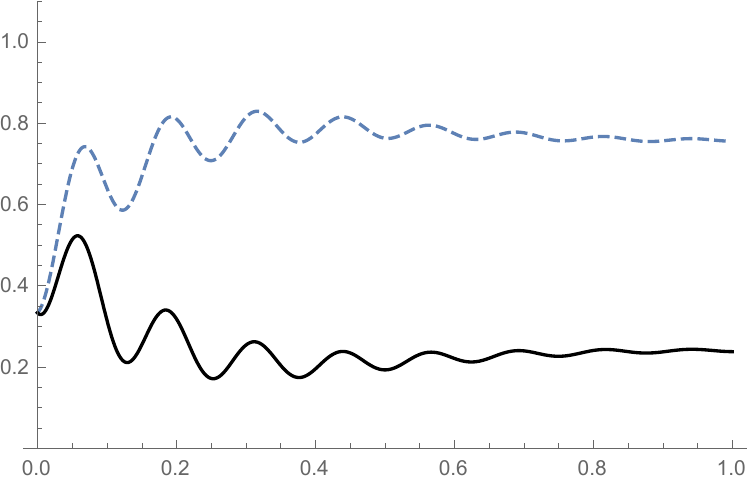}%
	\end{center}
	\caption{{\protect\footnotesize  The QFs $n_{1}(t)$ (dotted line) and $n_{2}(t)$
			(continuous line) for $\mu_{acm}=2$, $\mu_{cm}=25$, $N_1=1$, $N_2=0$ and the other parameters given in the text. Initial conditions: top left $\Psi_0^{(1)}$; top right $\Psi_0^{(2)}$; down left $\Psi_0^{(3)}$; down right $\Psi_0^{(4)}$.}}
	\label{fig15}
\end{figure}

\begin{figure}[th]
	\begin{center}
		\includegraphics[width=0.4\textwidth]{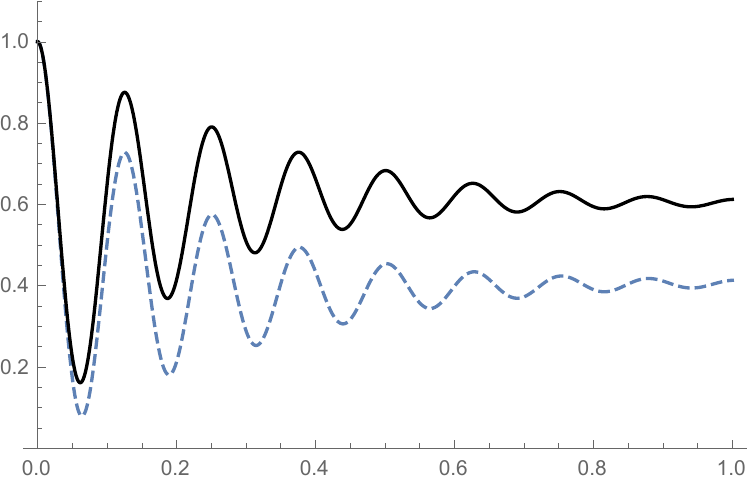}\hspace{%
			8mm} \includegraphics[width=0.4\textwidth]{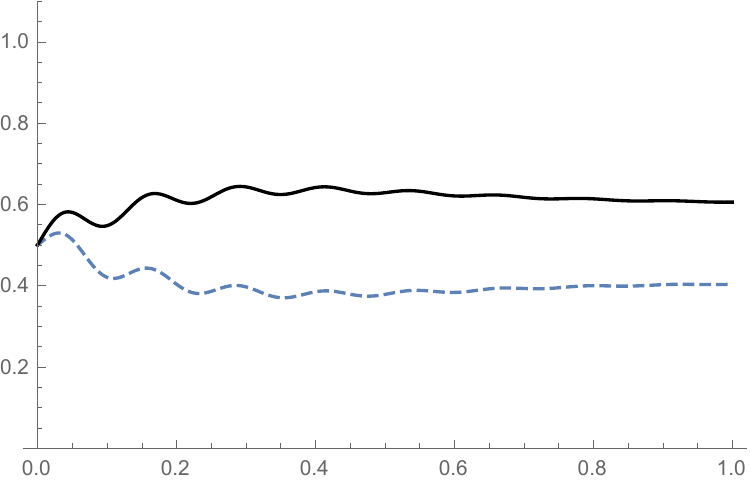}%
		\hfill\\[2pt]
		\includegraphics[width=0.4\textwidth]{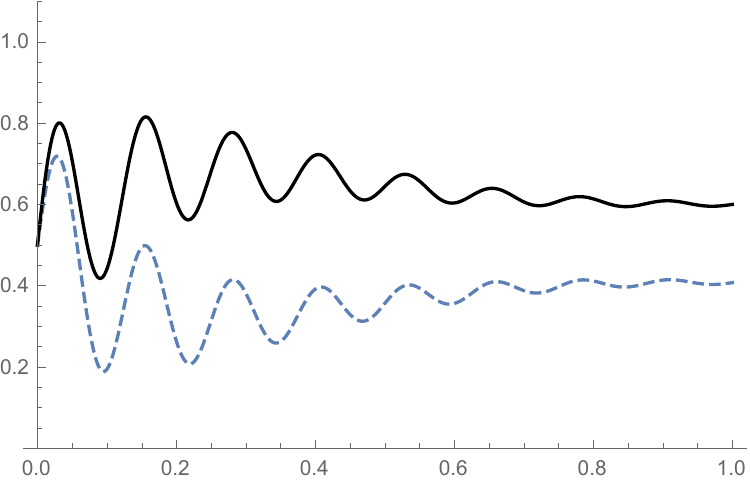}\hspace{%
			8mm} \includegraphics[width=0.4\textwidth]{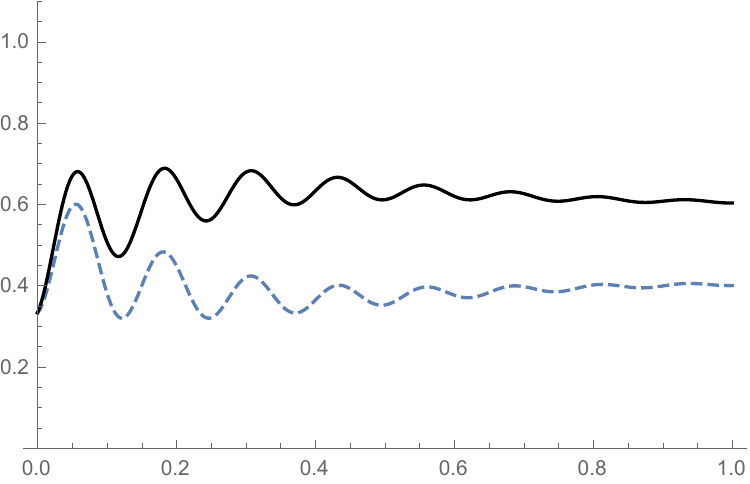}%
	\end{center}
	\caption{{\protect\footnotesize  The QFs $n_{1}(t)$ (dotted line) and $n_{2}(t)$
			(continuous line) for $\mu_{acm}=2$, $\mu_{cm}=25$, $N_1=1$, $N_2=1$ and the other parameters given in the text. Initial conditions: top left $\Psi_0^{(1)}$; top right $\Psi_0^{(2)}$; down left $\Psi_0^{(3)}$; down right $\Psi_0^{(4)}$.}}
	\label{fig16}
\end{figure}

\begin{figure}[th]
	\begin{center}
		\includegraphics[width=0.4\textwidth]{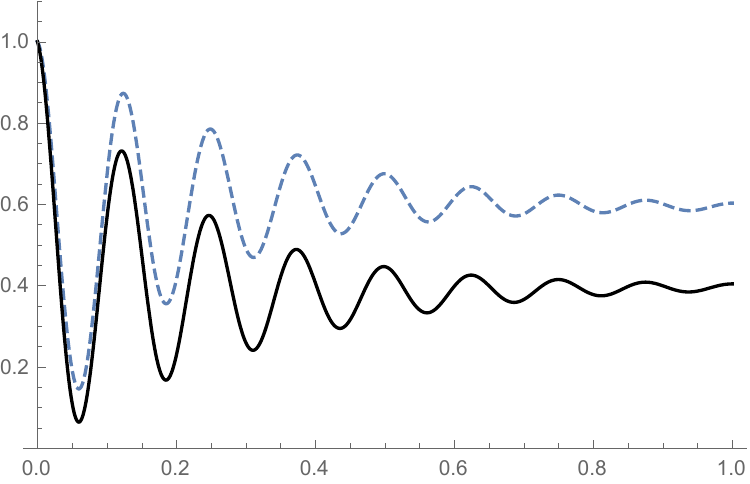}\hspace{%
			8mm} \includegraphics[width=0.4\textwidth]{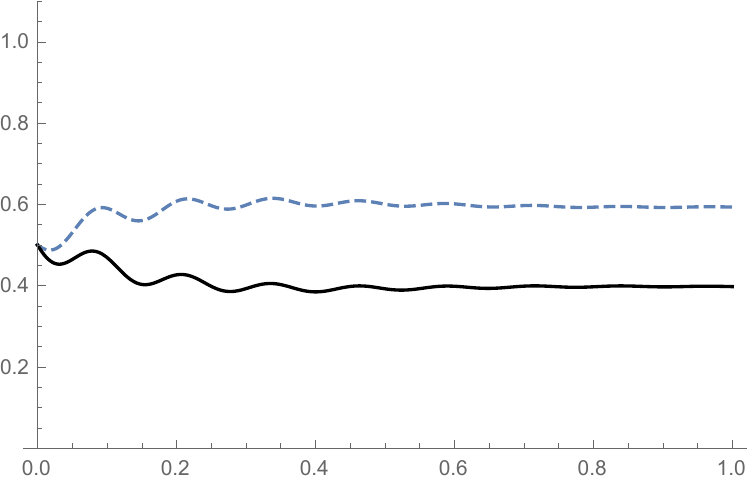}%
		\hfill\\[2pt]
		\includegraphics[width=0.4\textwidth]{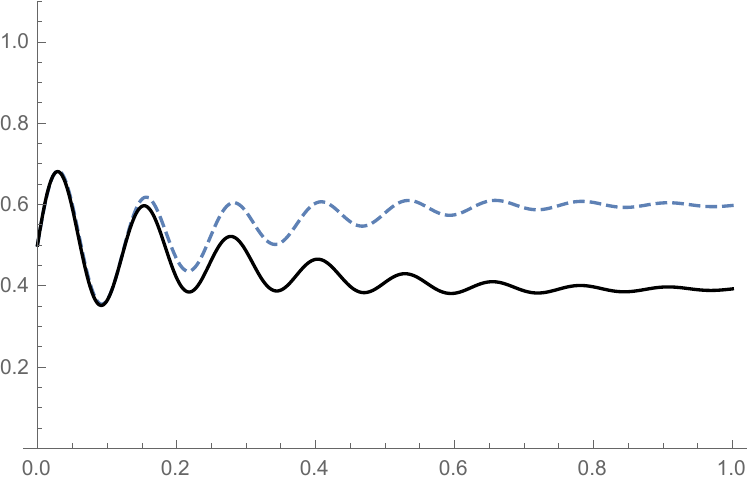}\hspace{%
			8mm} \includegraphics[width=0.4\textwidth]{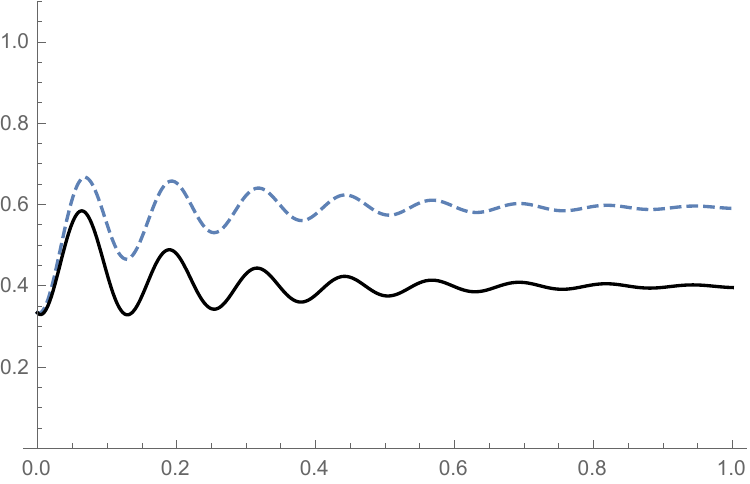}%
	\end{center}
	\caption{{\protect\footnotesize  The QFs $n_{1}(t)$ (dotted line) and $n_{2}(t)$
			(continuous line) for $\mu_{acm}=2$, $\mu_{cm}=25$, $N_1=0$, $N_2=0$ and the other parameters given in the text. Initial conditions: top left $\Psi_0^{(1)}$; top right $\Psi_0^{(2)}$; down left $\Psi_0^{(3)}$; down right $\Psi_0^{(4)}$.}}
	\label{fig17}
\end{figure}

\begin{figure}[th]
	\begin{center}
		\includegraphics[width=0.4\textwidth]{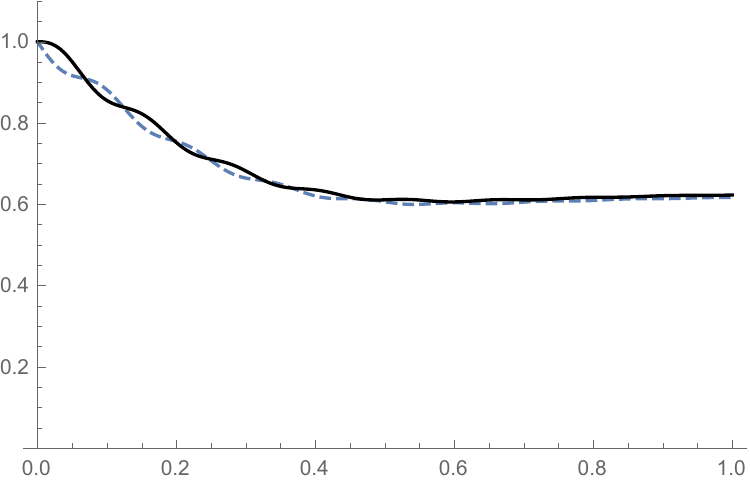}\hspace{%
			8mm} \includegraphics[width=0.4\textwidth]{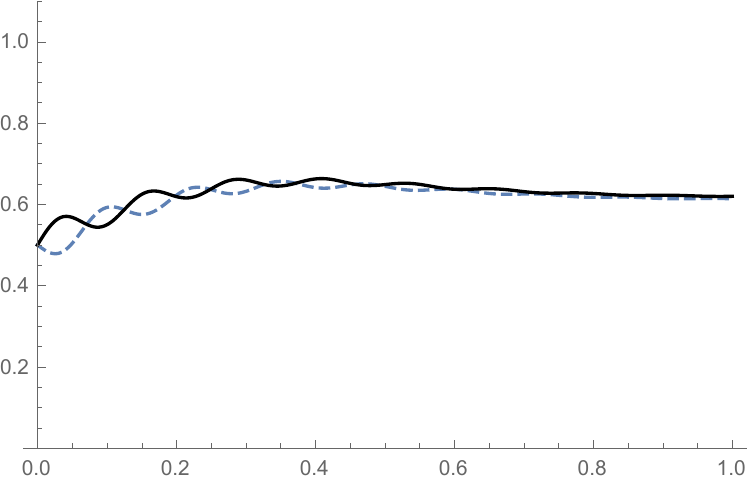}%
		\hfill\\[2pt]
		\includegraphics[width=0.4\textwidth]{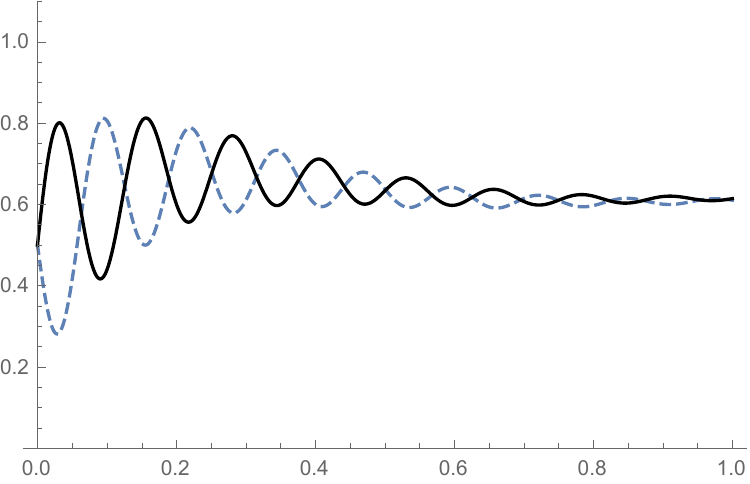}\hspace{%
			8mm} \includegraphics[width=0.4\textwidth]{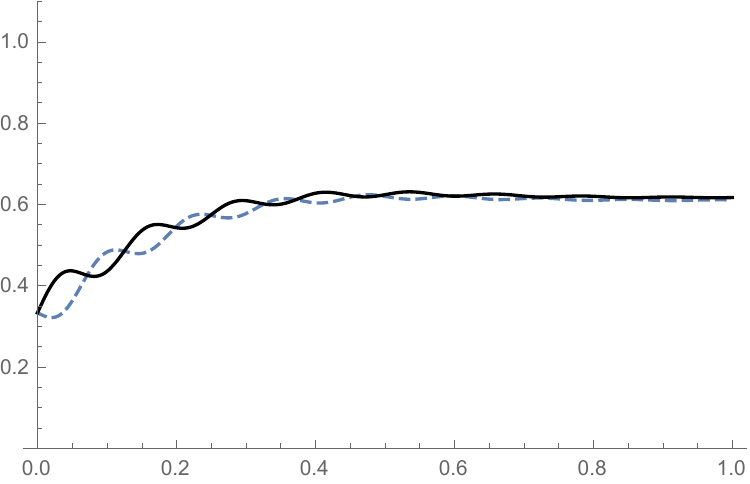}%
	\end{center}
	\caption{{\protect\footnotesize The QFs $n_{1}(t)$ (dotted line) and $n_{2}(t)$
			(continuous line) for $\mu_{acm}=25$, $\mu_{cm}=2$, $N_1=0$, $N_2=1$ and the other parameters given in the text. Initial conditions: top left $\Psi_0^{(1)}$; top right $\Psi_0^{(2)}$; down left $\Psi_0^{(3)}$; down right $\Psi_0^{(4)}$.}}
	\label{fig18}
\end{figure}

\begin{figure}[th]
	\begin{center}
		\includegraphics[width=0.4\textwidth]{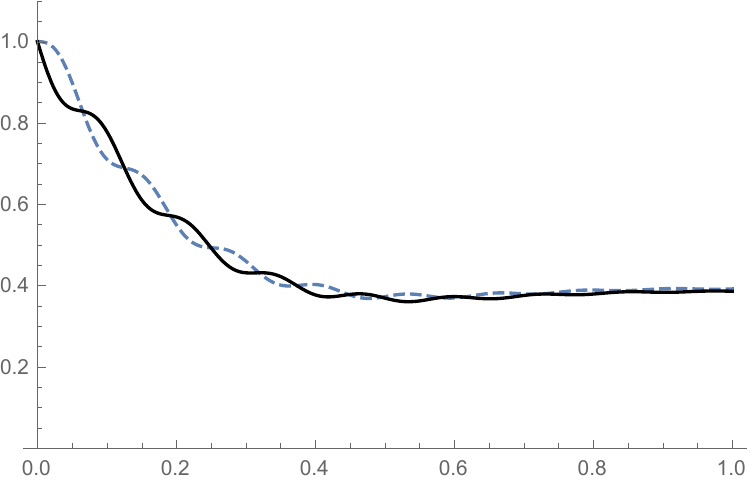}\hspace{%
			8mm} \includegraphics[width=0.4\textwidth]{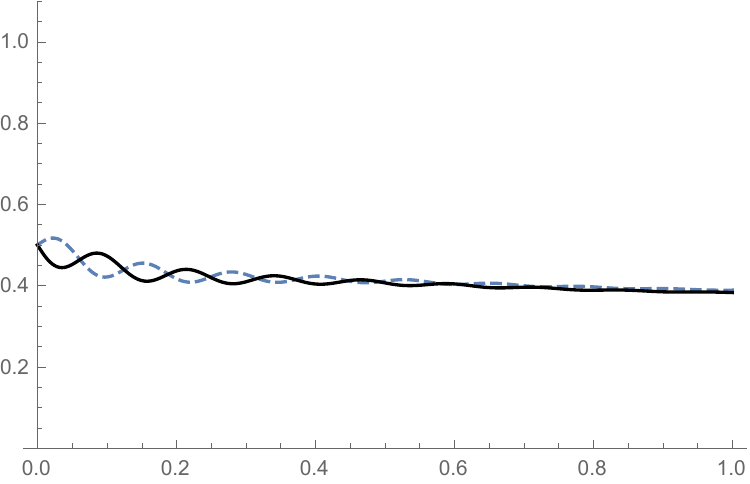}%
		\hfill\\[2pt]
		\includegraphics[width=0.4\textwidth]{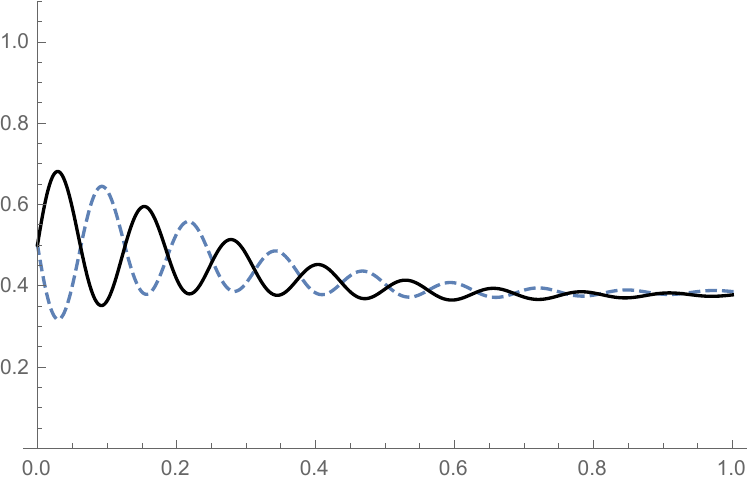}\hspace{%
			8mm} \includegraphics[width=0.4\textwidth]{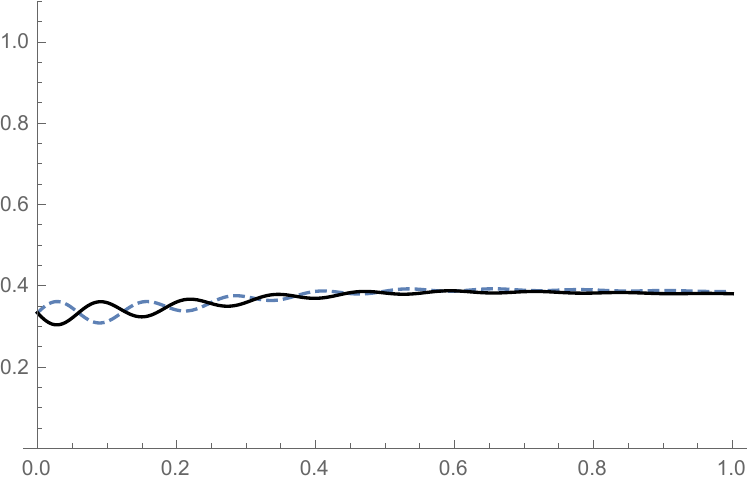}%
	\end{center}
	\caption{{\protect\footnotesize  The QFs $n_{1}(t)$ (dotted line) and $n_{2}(t)$
			(continuous line) for $\mu_{acm}=25$, $\mu_{cm}=2$, $N_1=1$, $N_2=0$ and the other parameters given in the text. Initial conditions: top left $\Psi_0^{(1)}$; top right $\Psi_0^{(2)}$; down left $\Psi_0^{(3)}$; down right $\Psi_0^{(4)}$.}}
	\label{fig19}
\end{figure}

\begin{figure}[th]
	\begin{center}
		\includegraphics[width=0.4\textwidth]{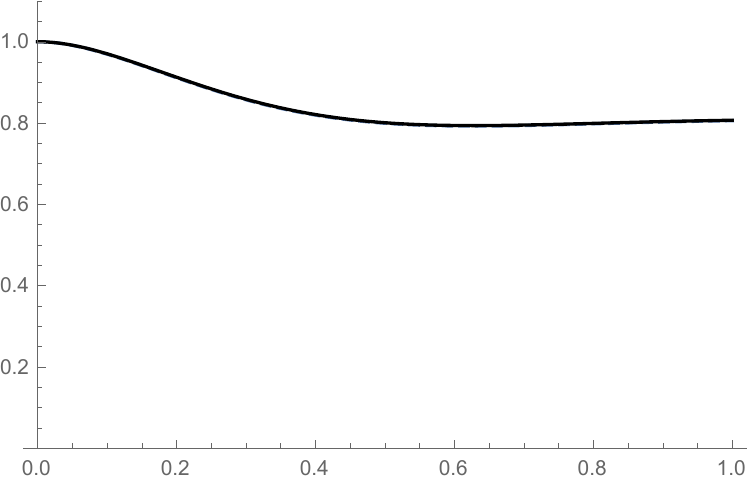}\hspace{%
			8mm} \includegraphics[width=0.4\textwidth]{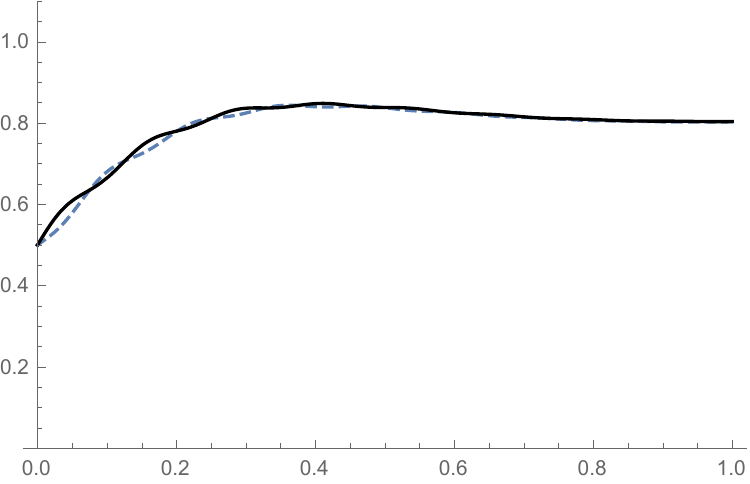}%
		\hfill\\[2pt]
		\includegraphics[width=0.4\textwidth]{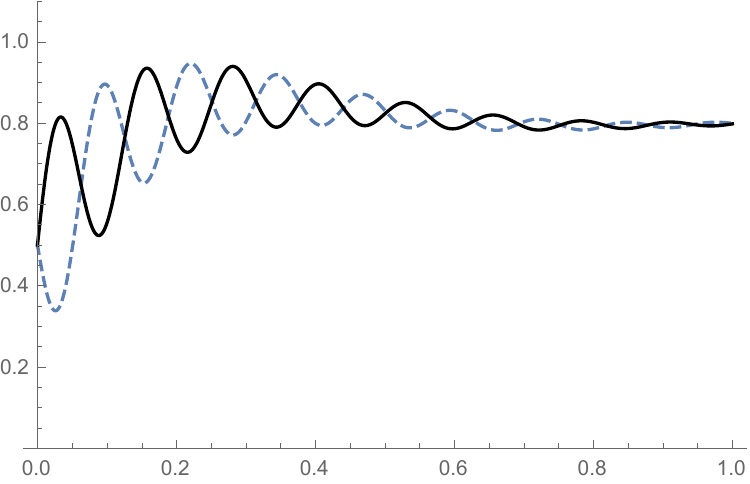}\hspace{%
			8mm} \includegraphics[width=0.4\textwidth]{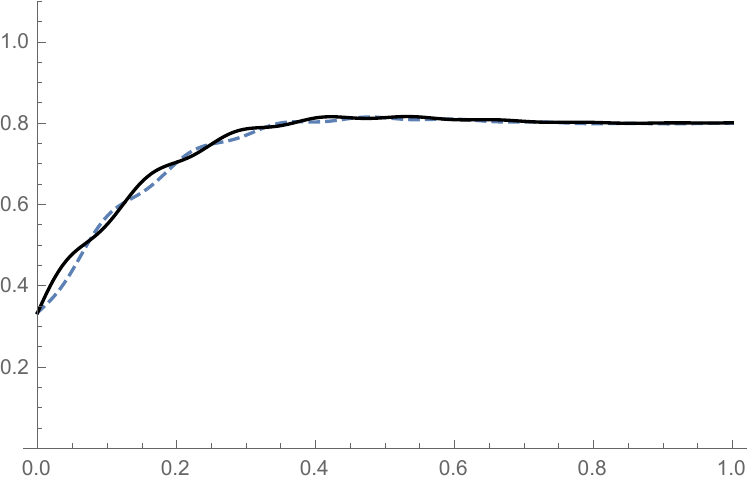}%
	\end{center}
	\caption{{\protect\footnotesize  The QFs $n_{1}(t)$ (dotted line) and $n_{2}(t)$
			(continuous line) for $\mu_{acm}=25$, $\mu_{cm}=2$, $N_1=1$, $N_2=1$ and the other parameters given in the text. Initial conditions: top left $\Psi_0^{(1)}$; top right $\Psi_0^{(2)}$; down left $\Psi_0^{(3)}$; down right $\Psi_0^{(4)}$.}}
	\label{fig20}
\end{figure}

\begin{figure}[th]
	\begin{center}
		\includegraphics[width=0.4\textwidth]{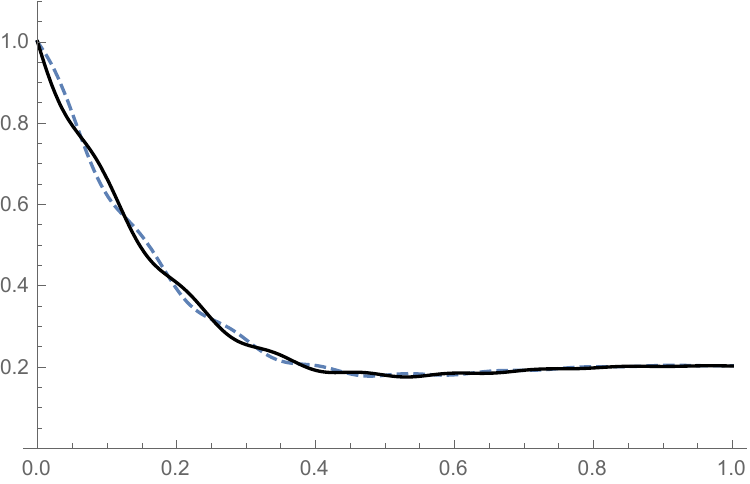}\hspace{%
			8mm} \includegraphics[width=0.4\textwidth]{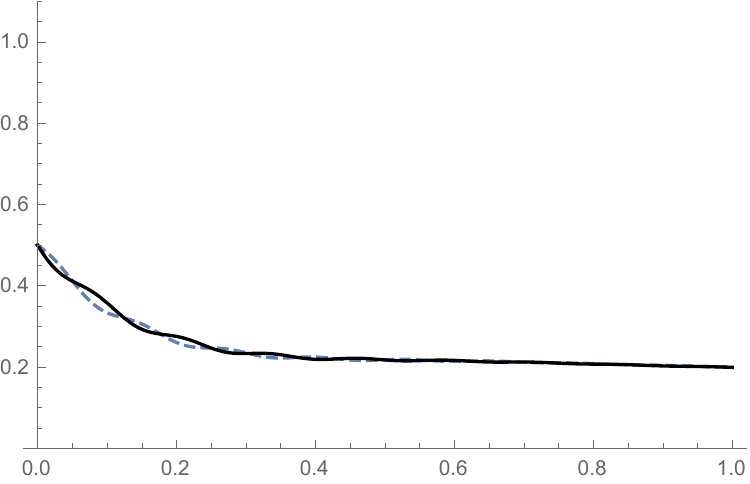}%
		\hfill\\[2pt]
		\includegraphics[width=0.4\textwidth]{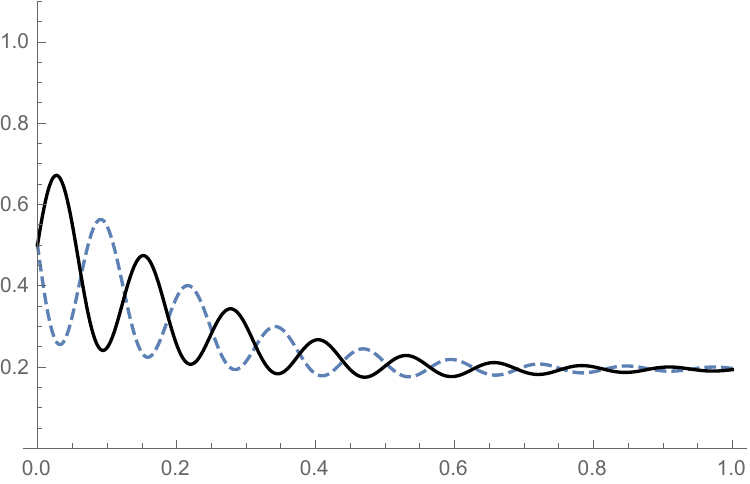}\hspace{%
			8mm} \includegraphics[width=0.4\textwidth]{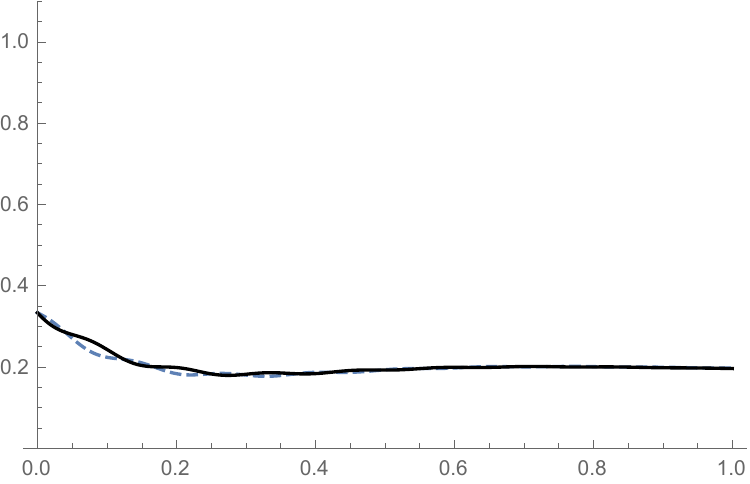}%
	\end{center}
	\caption{{\protect\footnotesize  The QFs $n_{1}(t)$ (dotted line) and $n_{2}(t)$
			(continuous line) for $\mu_{acm}=2$, $\mu_{cm}=25$, $N_1=0$, $N_2=0$ and the other parameters given in the text. Initial conditions: top left $\Psi_0^{(1)}$; top right $\Psi_0^{(2)}$; down left $\Psi_0^{(3)}$; down right $\Psi_0^{(4)}$.}}
	\label{fig21}
\end{figure}

Figures \ref{fig10}-\ref{fig13} are all based  on the following choice of $\mu$'s: $\mu_{acm}=2$, $\mu_{cm}=10$. These parameters change on \ref{fig14}-\ref{fig17}, where we have $\mu_{acm}=2$, $\mu_{cm}=25$. In Figures \ref{fig18}-\ref{fig21}, we have $\mu_{acm}=25$ and $\mu_{cm}=2$. 

From these figures we can draw the following conclusions:

\begin{enumerate}

	\item Despite of what happens when $\mu_{acm}=\mu_{cm}=0$, the asymptotic values of the QFs do not coincide with $N_1$ and $N_2$, at least for the time scales considered now. These values, contrarily to what was observed in Figure \ref{fig7}, are never zero or one, showing that divergent bank behaviors prevent the debt-/equity deposit mix of individual banks from taking extreme values.
	
	\item Still we observe in all plots in Figures \ref{fig10}-\ref{fig21} that the asymptotic value of $n_j(t)$ tends to stay closer to $N_j$, $j=1,2$, independently of the initial conditions (i.e., of the vector $\Psi_0^{(k)}$, $k=1,2,3,4$). This effect is particularly evident when $N_1\neq N_2$, again a sign of greater stability induced by divergent bank behaviors.

	\item Comparing the top right and the bottom left plots in each of the Figures \ref{fig10}-\ref{fig13}, we observe a serious difference in the width of the oscillations of the QFs. Since these plots correspond essentially to the same initial conditions ($n_1(0)=n_2(0)=\frac{1}{2}$), this might look strange. In fact, this effect was already observed previously, \cite{Baghavkhr}, and it is due to the presence of the strong interference terms in $n_j(t)$, due to the complex values of some of the $\alpha_{k,l}$: complex coefficients in the definition of $\Psi_0$ create stronger oscillations.

	\item In some of the plots we see that, during their time evolution, one of the QFs approaches the zero value. This is what happens, for instance, for $n_1(t)$ in Figure \ref{fig10}, top left and bottom left, as well as for $n_2(t)$ in Figure \ref{fig11}, top left. Similar effects are also found in Figures \ref{fig14} and \ref{fig15}, but not in the others. So, we have situations in which a bank, $\B_1$ or $\B_2$, depending on the values of $N_1$ and $N_2$, is extremely efficient, i.e., it reaches a zero value for $r_j^D$ (that, is, zero debt deposits over total deposits). Yet, this situation is not sustainable in a world of interacting banks (however large their EoS are) and doesn't last long; so the value of the QF increases again. What is interesting is that, as can be seen from Figure \ref{fig10}, we have $n_1(t_0)=0$ both if $n_1(0)=1$ and if $n_1(0)=\frac{1}{2}$. This feature seems to be not very much related to the initial condition of the bank.

	\item We observe that the width of the oscillations of the QFs is much smaller in Figures \ref{fig18}-\ref{fig21} than in the previous ones. This suggests that the effect of $\mu_{cm}$ in destabilizing the banks is much stronger than that of $\mu_{acm}$, as already observed in the second scenario: when the two banks behave similarly, their QFs oscillates more.
	
	\item While in Figures \ref{fig10} -\ref{fig17} the two QFs $n_1(t)$ and $n_2(t)$ are different one from the other, in Figures \ref{fig18} -\ref{fig21}, after some transient, they become almost indistinguishable: not only a larger $\mu_{acm}$ creates {\em small} oscillations, but it also makes the long-time behavior of the two banks similar.

\end{enumerate}

In conclusion, our model suggests that the most efficient
 (i.e., with lower long time value of $n_j(t)$) and stable (i.e., with smaller variations) system configuration is achieved for larger values of the ratio $\frac{\mu_{acm}}{\mu_{cm}}$, and for the value $N_1=N_2=0$ of the environments. In this case {\em the whole system is efficient}, and there is no difference between $\B_1$ and $\B_2$, except in a short transient. 

However, as we have already commented, there are other choices of the parameters which make one bank more efficient than the other. One such a choice is described in Figure \ref{fig7}, plots in the left columns, where the QFs are monotonically convergent to some asymptotic value. In all these cases $\B_1$ is much more efficient than $\B_2$. However, these plots are deduced in absence of interactions of any kind between the banks, and for this reason they are not realistic. More interesting is the situation described in Figure \ref{fig14}, top-right, where both $n_j(t)$ show very small oscillations, with a decreasing $n_1(t)$, which reaches a small value (in agreement with the fact that $N_1=0$), while $n_2(t)$ increases (again, in agreement with the fact that $N_2=1$): $\B_1$ is stable and improves its situation over time, while $\B_2$ makes its situation worse, even if it started from the same situation as$\B_1$ ($n_1(0)=n_2(0)=\frac{1}{2}$). A specular situation is shown in Figure \ref{fig15}, top-right, where the plots of $n_1(t)$ and $n_2(t)$ look exactly the opposite as those just mentioned in Figure \ref{fig14}. From this comparison we see the relevance of the environments in this aspect of the model: the smaller the value of the environment-determined $N_j$, the lower the asymptotic value of $n_j(t)$, and vice versa. In other words, the environment affects the banks and the system as a whole by strengthening or weakening their efficiency. We also see the relevance of the banks' interactions: the co-movement of their money creation (deposit issuance) raises their exposure to risks both by making their dynamics more volatile and by widening the gap between the bank's actual values of their debt/equity deposit mix and the optimal value. The reverse happens if banks' move their money creation in opposite directions.

\section{Conclusions and perspectives}

\label{sectconcl}

{According to the Accounting View of Money (AVM), and as discussed in Section I of this article, the money issued by commercial banks in the form of demand deposits features a hybrid nature, since deposits can be shown to consist of a share of deposits bearing the characteristics of debt (debt-deposits) and a share of deposits bearing the characteristics of equity (equity-deposits), in a mix that depends on factors that relate to the issuing banks and the environment where they operate and interact, which may change over time.
Following this important finding of the AVM, it was only consequential to associate the hybrid nature of bank deposits with the dual nature of the objects of quantum physics, and to investigate whether and how the application of quantum analytical methods to a form of money showing dualistic features could be used to extract valuable economic information.
This article has thus studied demand deposits (which represents the prevailing form of money in all contemporary economies) as a possible form of “quantum money," and has analyzed it through methods widely adopted in quantum mechanics, using a quantum model to describe some relevant aspects of it, including for instance how banks’ power to create money (i.e., issue deposits typically via the credit channel) is affected by the interactions taking place between the banks and between the banks and their environments.
The main general conclusion that can be drawn from the preliminary use of the quantum money model presented in this study, and from the analytical results and numerical simulations, is that efficient and stable banking systems are highly diversified, that is, they are populated by banks of different size, operating in different environments, and behaving differently. Such systems are more efficient and stable than less diversified ones.
Here, the terms “efficiency” and “stability” refer to values of $n_j(t)$ that are low and not volatile, over time, and reflect banking systems that are characterized by high capacity to create money and limited exposure to risks; that is, their debt/equity deposit mix is (close to) optimal and does not expose the banks to significant liquidity, credit and settlement risks.
Another important result of the model is that systems where all banks co-move, that is, change their deposit-issuance behavior in the same direction (i.e., either increasing or decreasing it) are less stable than systems where banks change it in opposite directions, since risks in the latter are mitigated.}
{This article has laid the foundations of the study of the dual nature of bank money - conjectured by the AVM - and its association with a quantum object to study the effects on bank money creation of the interactions that take place between banks and between the banks and the environment where they operate. The results obtained through the application of quantum mechanics to bank money are only preliminary and suggest that important knowledge gains can be attained through this new approach in terms of system behaviors and properties. 
As we mentioned earlier, we have been initially motivated by our interest in setting the foundations of this new approach, and we used it to investigate some qualitative properties of a very simple, stylized banking system. For this reason, we were not particularly interested in considering specific experimental data. However, looking forward, this is quite a relevant step, and as part of our future plans we intend to improve the model proposed in this article here to fit experimental data. We believe that the use of the $(H,\rho)$-induced dynamics, \cite{bagrules}, can play a relevant role in this extension since, as discussed in many other concrete situations, it is a realistic way to introduce in the analysis of financial (but not only) systems the effects of phenomena that are not easy to describe with a pure Hamiltonian framework, and which may have truly significant financial implications, such as wars, crises, sudden shortages of critical resources, etc.).

This is thus just the first step towards a better understanding of the relevance of quantum mechanical tools and ideas in connection with economics and finance. More work is in progress. 

The advantage is that most of the economic variables that are critical for the quantum analysis of bank deposits (as reported in Appendix A) are observables that would be relatively easy to proxy and measure, and to use in simulation scenarios, including: money supply (central bank reserves and commercial bank deposits); the interaction among banks and between banks and their environments; the factors affecting the environments; the probability of banks changing deposit and lending behavior (and more broadly, their assets and liabilities management; the money creation power of different banks; and the marginal cost involved in banks changing their money creation. Of course, new variables might become necessary, depending on the future evolution of the quantum analysis of bank deposits.
}

\section*{Data accessibility statement}

This work does not have any experimental data.

\section*{Competing interests statement}

We have no competing interests.

\section*{Authors' contributions}

BB introduced the concept of bank money as a quantum object, FB proposed the quantum money model and worked out its solutions, and BB interpreted the model's results in economic terms.

\section*{Acknowledgements}

One of the authors (FB) acknowledges partial support from the University of
Palermo and from GNFM of the INDAM. The authors wish to acknowledge the very helpful comments and suggestions of the anonymous referees of QEF, which have certainly contributed to improving the contents and the quality and the clearness of our study.

\section*{Funding statement}

This work received no financial support.

\renewcommand{\theequation}{A.\arabic{equation}}

\section*{Appendix A: The Model’s Variables and Parameters – Synoptic Table}\label{appendixA}

\begin{figure}[th]
	\begin{center}
		\includegraphics[width=0.95\textwidth]{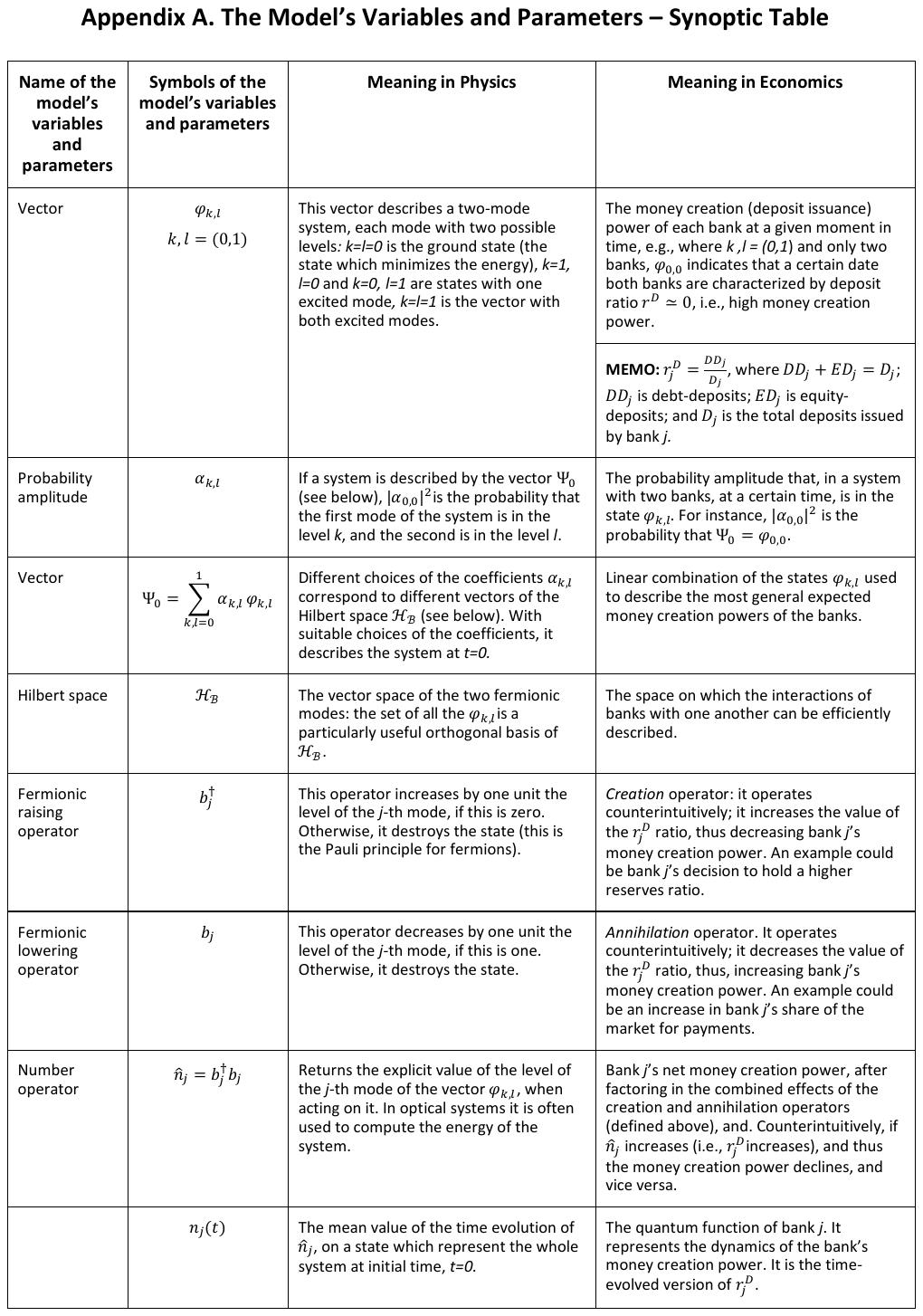}
	\end{center}
	\caption{{\protect\footnotesize The Synoptic Table, part 1 }}
	\label{table1}
\end{figure}

\begin{figure}[th]
	\begin{center}
		\includegraphics[width=0.9\textwidth]{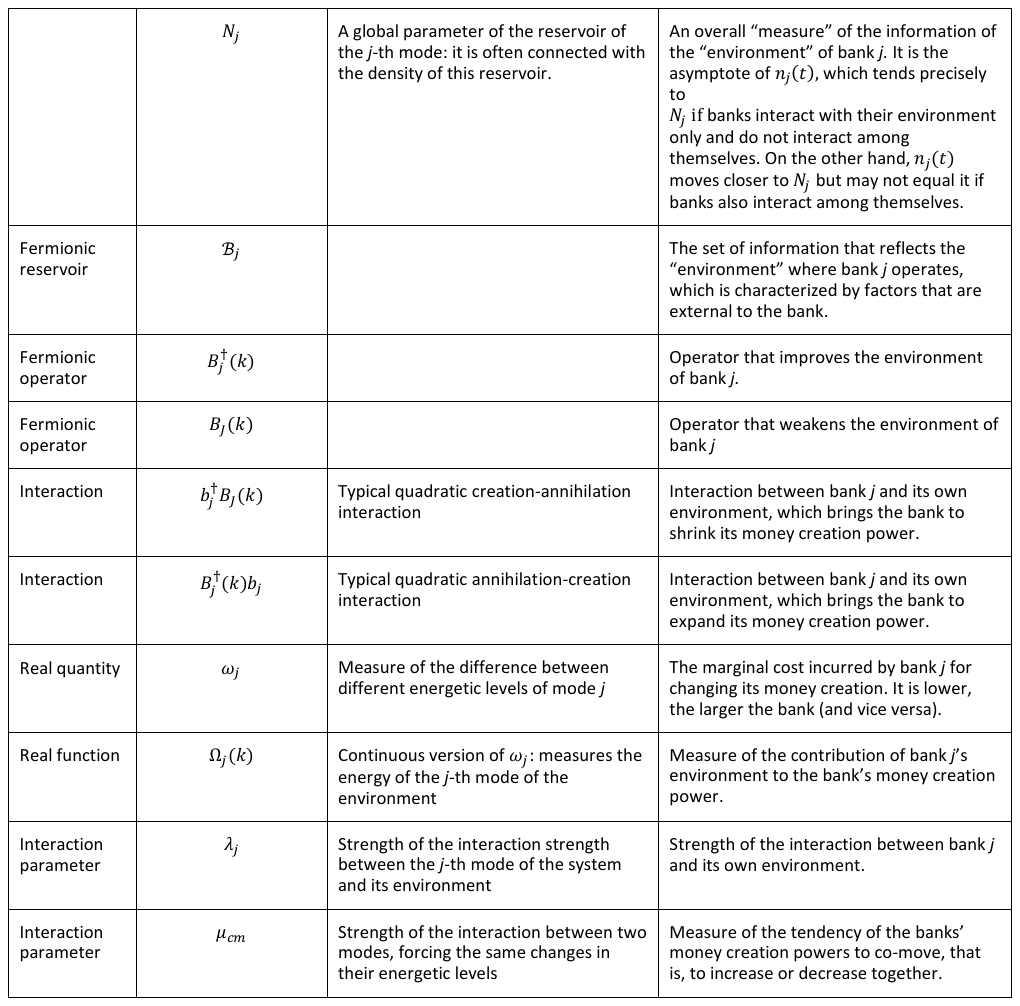}
	\end{center}
	\caption{{\protect\footnotesize The Synoptic Table, part 2 }}
	\label{table2}
\end{figure}

\begin{figure}[th]
	\begin{center}
		\includegraphics[width=0.9\textwidth]{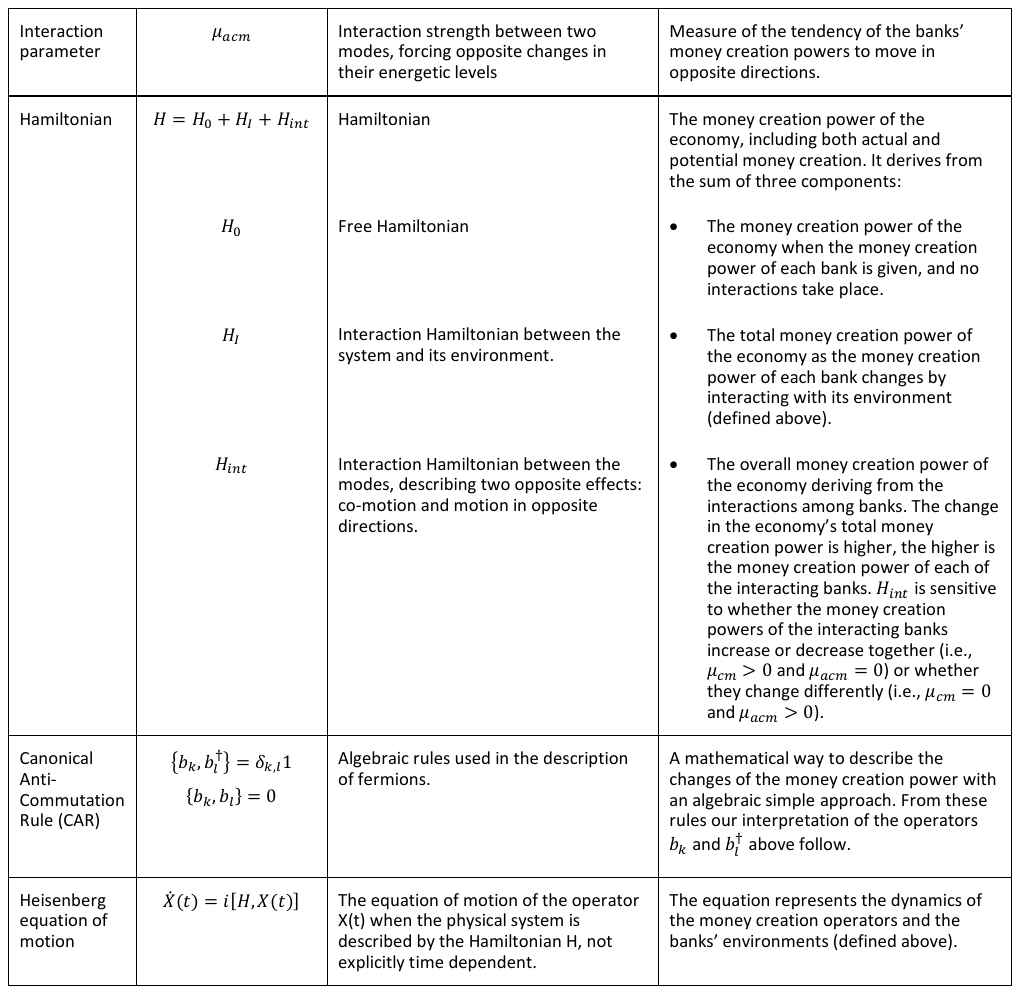}
	\end{center}
	\caption{{\protect\footnotesize The Synoptic Table, part 3 }}
	\label{table3}
\end{figure}

In this Appendix we propose a sort of vocabulary to translate and relate the quantum mechanical and enonomical meaning of the relevant quantities (parameters, operators, variables) used in this paper. This vocabulary is divided in three parts, Figure \ref{table1}, Figure \ref{table2} and Figure \ref{table3}.

\renewcommand{\theequation}{B.\arabic{equation}}

\section*{Appendix B: further plots}\label{appendixB}

\label{sectnumres}

In this Appendix we want to show a very special feature of our model, which was already observed in other contexts before, \cite{Baghavkhr}. Among the other parameters of the model, the coefficients $\alpha_{k,l}$ appearing in (\ref{20}) are important. In fact, as we have already discussed in details, they describe the initial status of the two banks. The normalization condition $\sum_{k,l=0}^{1}|\alpha _{k,l}|^{2}=1$ suggests us the use of complex quantities. In this perspective, it is interesting to see that serious differences arise if we take real or complex $\alpha_{k,l}$. To show this aspect of the model,  we start introducing a set  $\mathcal{C}$ corresponding to  the
following choice of parameters of $H$: $\omega _{1}=1$, $\omega _{2}=2$, $%
\Omega _{1}=\Omega _{2}=0.1$, $\lambda _{1}=\lambda _{2}=0.5$. We will also use a second choice, and $\mathcal{%
	C}_{2}$: $\omega _{1}=1.2$, $\omega _{2}=3$, $\Omega
_{1}=\Omega _{2}=0.3$, $\lambda _{1}=1=\lambda _{2}=0.4$. Also, we call $\alpha_{\mathbb{R}}$ and $\alpha_{\mathbb{C}}$ the following choices
of the parameters $\alpha _{k,l}$ in (\ref{20}): $\mathcal{C}_{\alpha,1
}=\{\alpha _{k,l}=\frac{1}{2},\forall \,k,l\}$, while $\mathcal{C}_{\alpha,2
}=\{\alpha _{0,1}=\frac{1}{2}=-\alpha _{1,1},\,\alpha _{0,0}=\frac{i}{2}%
=-\alpha _{1,0}\}$.

\begin{figure}[th]
	\begin{center}
		\includegraphics[width=0.4\textwidth]{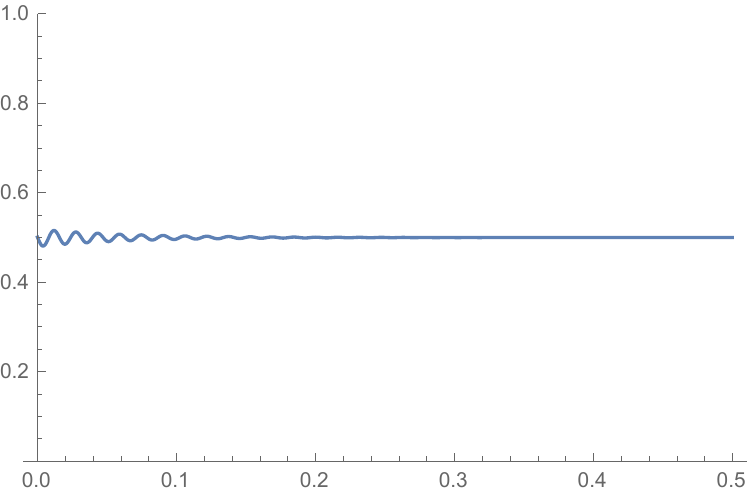}\hspace{%
			8mm} \includegraphics[width=0.4\textwidth]{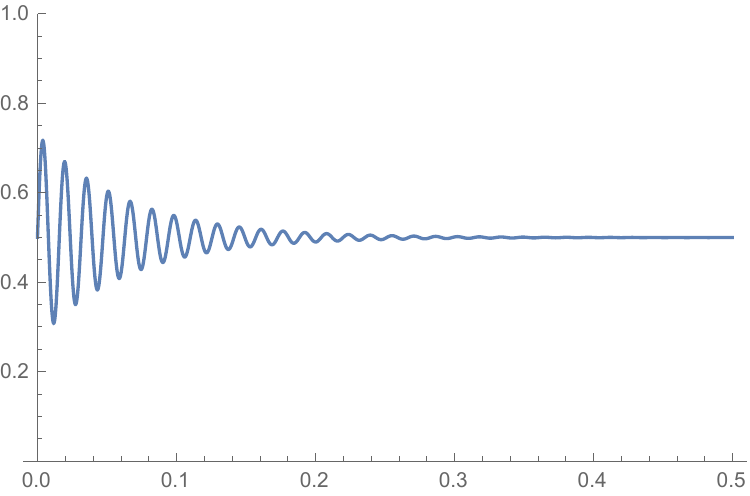}%
		\hfill\\[0pt]
		\includegraphics[width=0.4\textwidth]{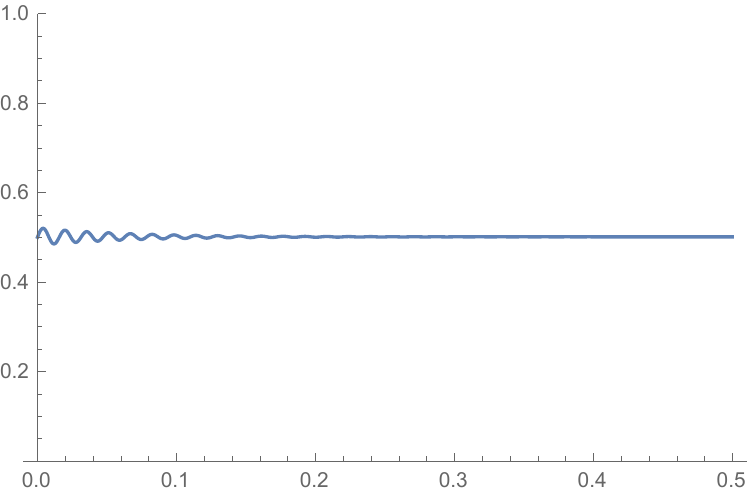}\hspace{%
			8mm} \includegraphics[width=0.4\textwidth]{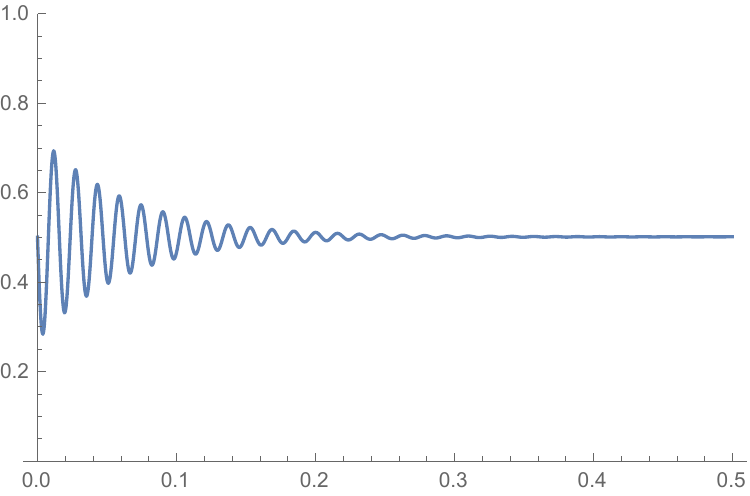}%
		\hfill\\[0pt]
	\end{center}
	\caption{{\protect\footnotesize The QFs $n_{1}(t)$ (up) and $n_{2}(t)$
			(down) for parameters $\mathcal{C}_{1}$, $N_{1}=0$, $N_{2}=1$, $\protect\mu %
			_{acm}=200$, $\protect\mu_{cm}=0$ and for $\alpha_{\mathbb{R}}$
			(left) and $\alpha_{\mathbb{C}}$ (right). }}
	\label{fig1}
\end{figure}

\begin{figure}[th]
	\begin{center}
		\includegraphics[width=0.4\textwidth]{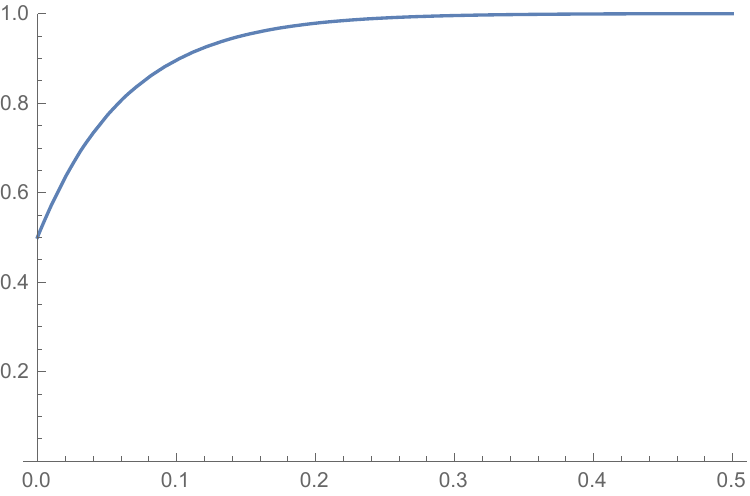}\hspace{%
			8mm} \includegraphics[width=0.4\textwidth]{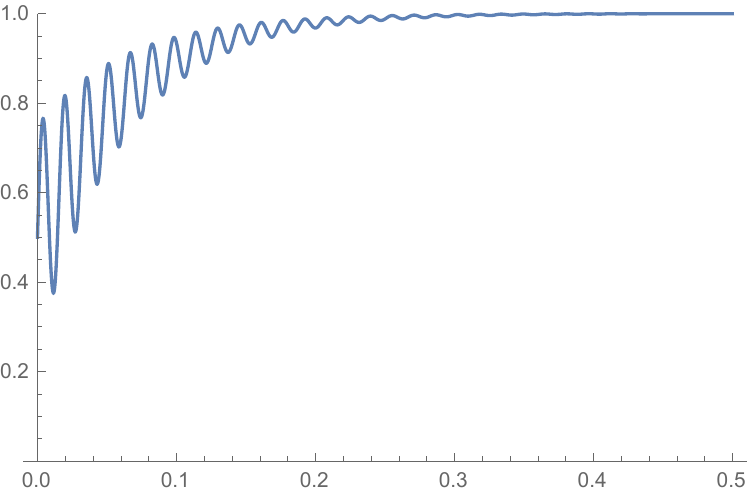}%
		\hfill\\[0pt]
		\includegraphics[width=0.4\textwidth]{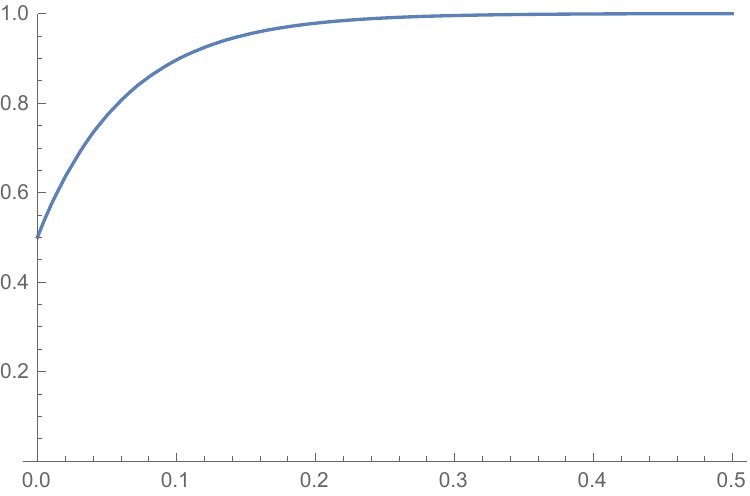}\hspace{%
			8mm} \includegraphics[width=0.4\textwidth]{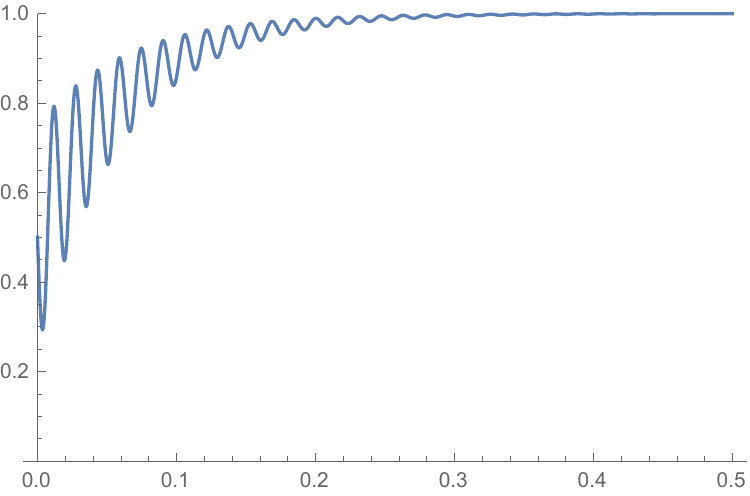}%
		\hfill\\[0pt]
	\end{center}
	\caption{{\protect\footnotesize The QFs $n_{1}(t)$ (up) and $n_{2}(t)$
			(down) for parameters $\mathcal{C}_{1}$, $N_{1}=1$, $N_{2}=1$, $\protect\mu %
			_{acm}=200$, $\protect\mu_{cm}=0$ and for $\alpha_{\mathbb{R}}$
			(left) and $\alpha_{\mathbb{C}}$ (right). }}
	\label{fig2}
\end{figure}

\begin{figure}[th]
	\begin{center}
		\includegraphics[width=0.4\textwidth]{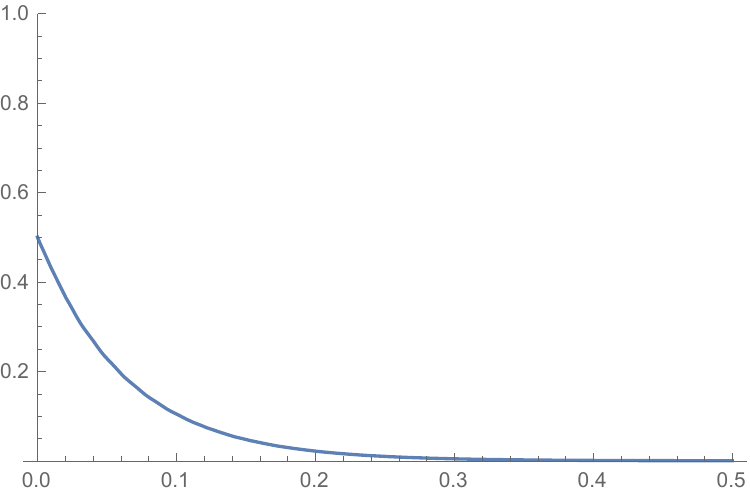}\hspace{%
			8mm} \includegraphics[width=0.4\textwidth]{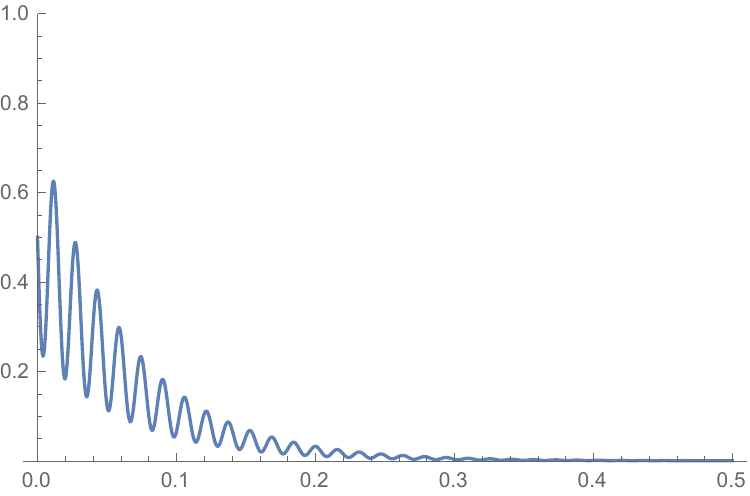}%
		\hfill\\[0pt]
		\includegraphics[width=0.4\textwidth]{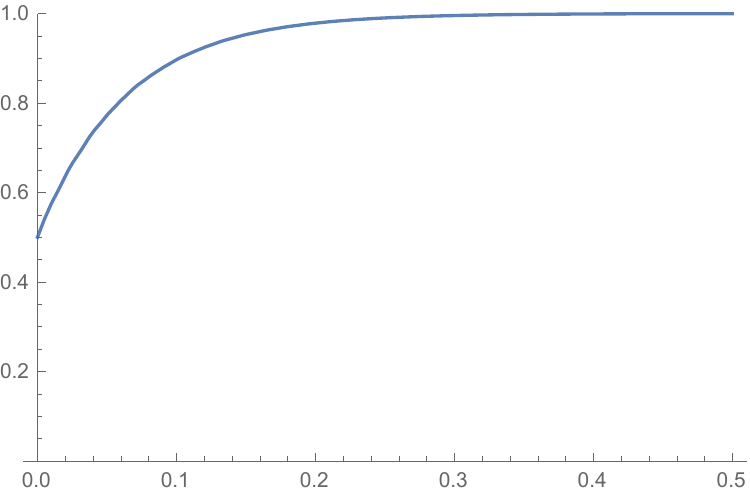}\hspace{%
			8mm} \includegraphics[width=0.4\textwidth]{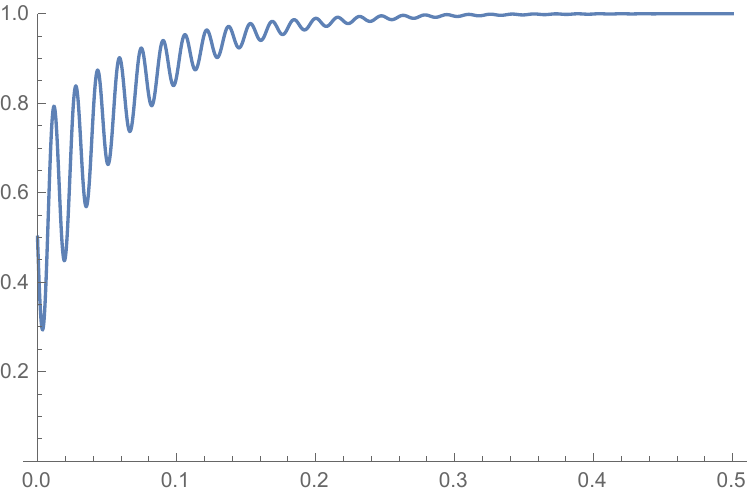}%
		\hfill\\[0pt]
	\end{center}
	\caption{{\protect\footnotesize The QFs $n_{1}(t)$ (up) and $n_{2}(t)$
			(down) for parameters $\mathcal{C}_{1}$, $N_{1}=0$, $N_{2}=1$, $\protect\mu %
			_{acm}=0$, $\protect\mu_{cm}=200$ and for $\alpha_{\mathbb{R}}$
			(left) and $\alpha_{\mathbb{C}}$ (right). }}
	\label{fig3}
\end{figure}

\begin{figure}[th]
	\begin{center}
		\includegraphics[width=0.4\textwidth]{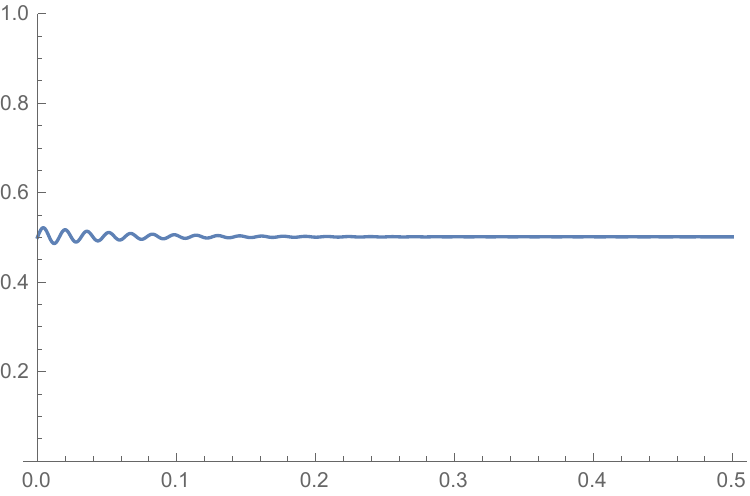}\hspace{%
			8mm} \includegraphics[width=0.4\textwidth]{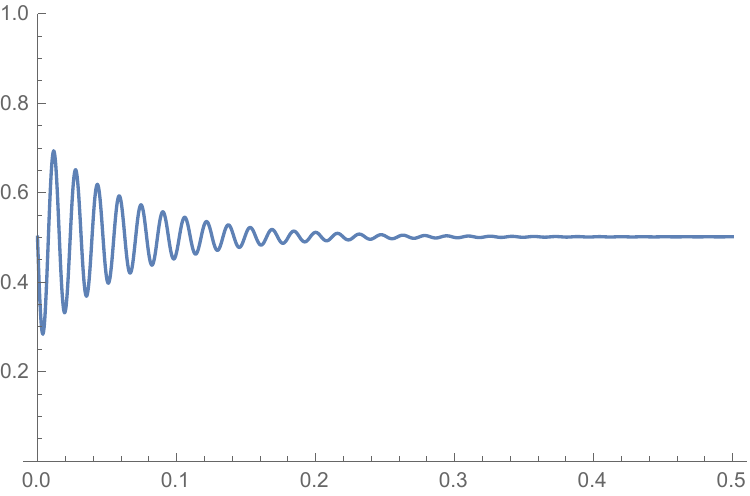}%
		\hfill\\[0pt]
		\includegraphics[width=0.4\textwidth]{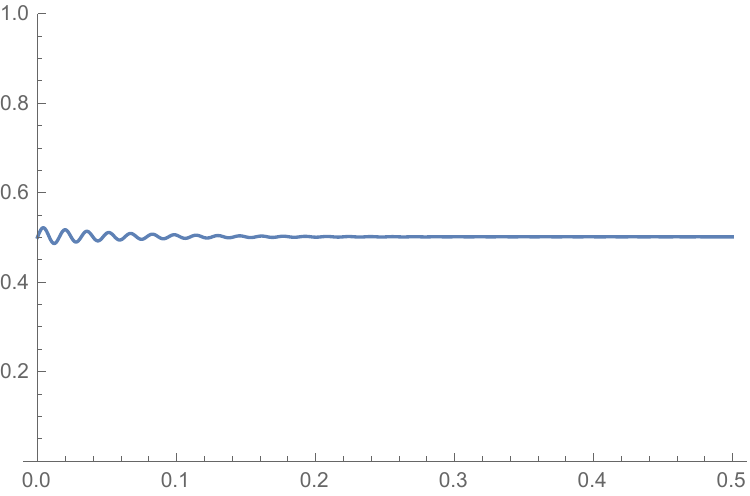}\hspace{%
			8mm} \includegraphics[width=0.4\textwidth]{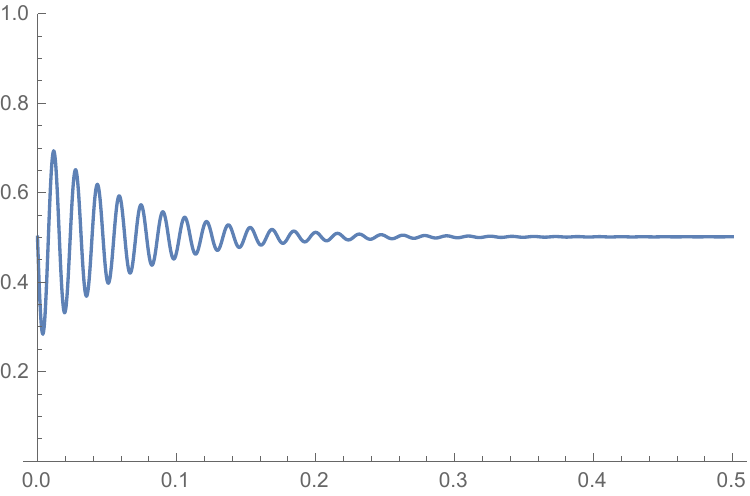}%
		\hfill\\[0pt]
	\end{center}
	\caption{{\protect\footnotesize The QFs $n_{1}(t)$ (up) and $n_{2}(t)$
			(down) for parameters $\mathcal{C}_{1}$, $N_{1}=1$, $N_{2}=1$, $\protect\mu %
			_{acm}=0$, $\protect\mu_{cm}=200$ and for $\alpha_{\mathbb{R}}$
			(left) and $\alpha_{\mathbb{C}}$ (right). }}
	\label{fig4}
\end{figure}

Several other choices could be considered, but this choice is sufficient to show what we are after, that is, the different kind of oscillations that can be observed during the time evolution of the interacting banks in the two cases. We anticipate that these choices do not affect the asymptotic values of the
two QFs, which appear to be independent of the choice of $\alpha_{\mathbb{R}}$
and $\alpha_{\mathbb{C}}$, even if there exists a certain time window in which
the choice of the $\alpha _{k,l}$'s really change the behaviors of the
functions. More concretely, if we add a phase in the coefficients defining
the original vector $\Psi _{0}$, we may observe quite large oscillations.
Then, \emph{interference terms in $\Psi _{0}$ make it, in general, the value of the QFs quite
	oscillatory}. In particular, this is the effect of the
relative phases in the interference coefficients (see Figures \ref{fig1}-\ref{fig4},
right), while if these coefficients have all the same phases, no evident oscillations are observed (see Figures \ref{fig1}-\ref{fig4}, left). However, if we wait for
a sufficiently long time, in both cases we reach the same final values of
the QFs: the asymptotic values of the QFs only depend on the state of the
environment, and not on the particular choice of $\Psi _{0}$. Moreover, the presence of complex coefficients in (\ref{20}) produce a {\em first-stage instability} of the system, with large oscillations which only lower down after some time. So these non trivial phases in the coefficients $\alpha_{k,l}$ could be useful to model, for instance, some extreme financial effect in the market.



\end{document}